\documentclass[12pt,article]{iopart}

\usepackage{epsfig}
\usepackage{amscd,amssymb}
\usepackage{graphicx,xspace}
\usepackage{grffile}
\usepackage{color}

\newcommand{\eqref}[1]{(\ref{#1})}
\def\tauqtyp{\tau_q^{\mbox{\scriptsize $\textrm{typ}$}}}
\def\tauq{\tau_q}

\def\ftyp{f^{\mbox{\scriptsize $\textrm{typ}$}}}

\begin{document}\newcommand \be  {\begin{equation}}
\newcommand \bea {\begin{eqnarray} \nonumber }
\newcommand \ee  {\end{equation}}
\newcommand \eea {\end{eqnarray}}

\title[High values of disorder-generated multifractals]{High values of disorder-generated multifractals and logarithmically correlated processes}

\author{Yan V Fyodorov}

\address{School of Mathematical Sciences, Queen Mary University of London\\  London E1 4NS, United Kingdom}

\author{Olivier Giraud}

\address{LPTMS, CNRS and Universit\'e Paris-Sud, UMR 8626, B\^at.~100, 91405 Orsay, France}

\begin{abstract}
  In the introductory section of the article we give a brief account of recent insights into statistics of high and extreme values of disorder-generated  multifractals following a recent work by the first author with P. Le Doussal and A. Rosso (FLR) employing a close relation between multifractality and logarithmically correlated random fields. We then substantiate some aspects of the FLR approach analytically for multifractal eigenvectors in the Ruijsenaars-Schneider ensemble (RSE) of random matrices introduced by E. Bogomolny and the second author by providing an {\it ab initio} calculation that reveals hidden logarithmic correlations at the background of the disorder-generated multifractality. In the rest we investigate numerically a few representative models of that class, including the study of the highest component of multifractal eigenvectors in the Ruijsenaars-Schneider ensemble.

\end{abstract}

\maketitle

\section{Introduction}

\subsection{General setting}
Multifractal patterns are patterns of intensities which are characterized by a high variability over a wide range of space or time scales, and by huge fluctuations which can be visually detected. They have been observed and investigated in many areas of science, from physics, chemistry, geophysics, oceanology \cite{multif1,multif2} to climate studies \cite{climate} or mathematical finance \cite{multifinance1,multifinance2}. The multifractal approach has also proved relevant in fields such as growth processes \cite{DLA}, turbulence \cite{turb,Schertzer}, and the theory of quantum disordered systems \cite{EM2008}.
\begin{figure}[h!]
\begin{center}
\includegraphics[width=0.5\textwidth,angle=-90]{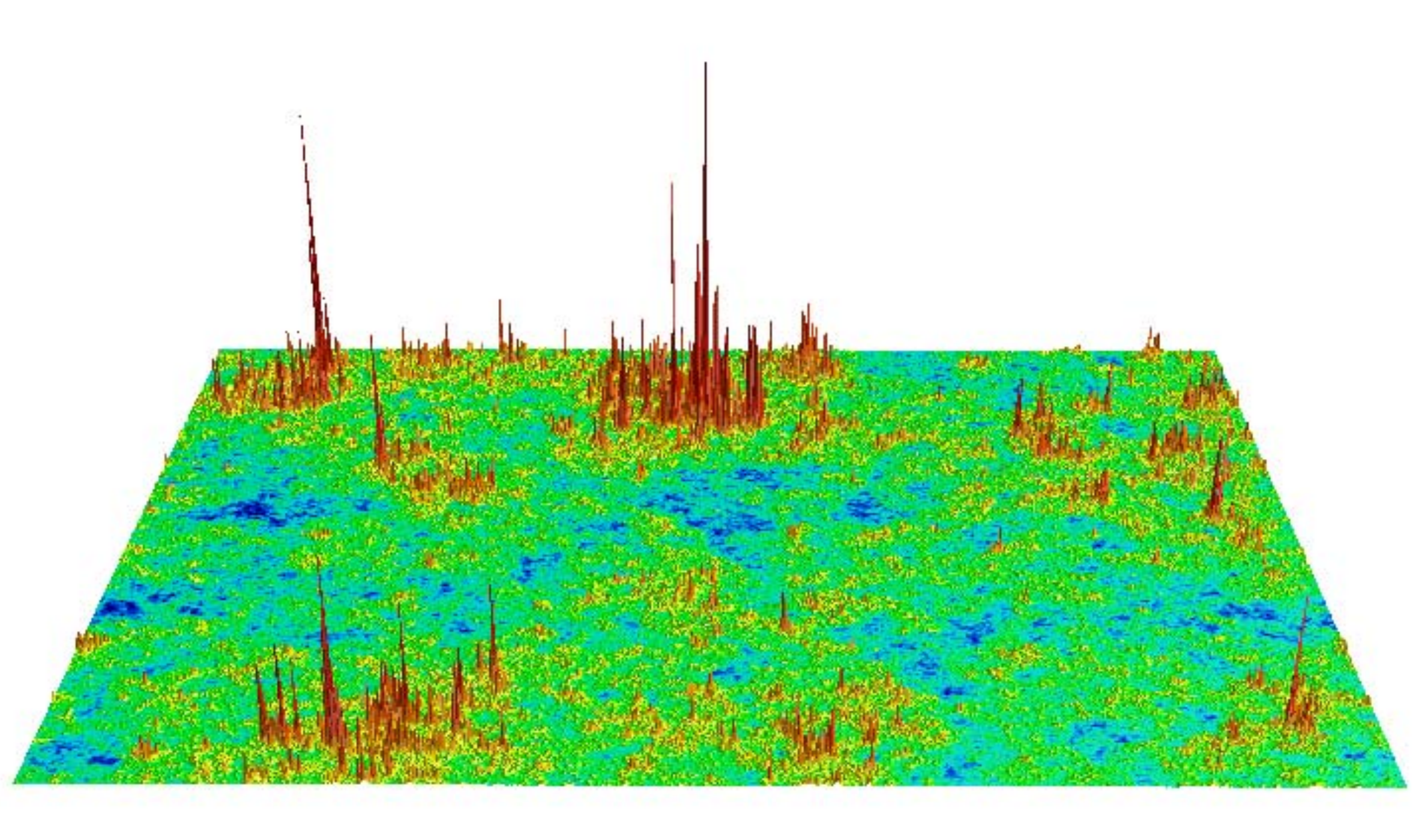}
\end{center}
\caption{ Intensity of a multifractal wavefunction at the point of Integer Quantum Hall Effect.
{\small  Courtesy of F. Evers,  A. Mirlin and A. Mildenberger. }}
\end{figure}

In a $d$-dimensional lattice of linear size $L$ and lattice spacing $a$, thus containing $M=(L/a)^d\gg 1$ lattice sites, multifractal patterns with intensities $h_i>0$ at different sites $i=1,\ldots M$ are characterized by attributing a different scaling $h_i\sim M^{x_i}$ to each intensity, with exponents $x_i$ forming a dense set. One of the most natural characteristics of a multifractal is the function ${\cal N}_M(x)$ counting the number of points in the pattern with exponents exceeding the value $x$. Introducing the density of exponents $\rho_M(x)$, so that ${\cal N}_M(x)=\int_x^{\infty}\,\rho_M(y)\,dy$, multifractality is equivalent to the statement that such a density behaves for $M\gg 1$ as
\begin{equation}\label{1}
\rho_M(x)=\sum_{i=1}^M\,\delta\left(\frac{\ln{h_i}}{\ln{M}}-x\right)\approx c_M(x) \sqrt{\ln{M}}\, M^{f(x)}, \quad M\gg 1,
\end{equation}
where $f(x)$, the singularity spectrum, is a function of $x$, and $c_M(x)$ is of order unity. This is frequently referred to as  the {\it multifractal Ansatz}.
 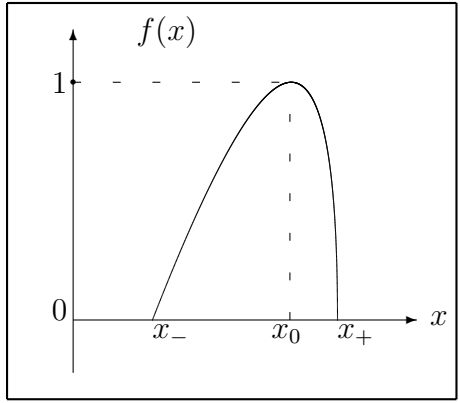
\begin{figure}[h]
\begin{picture}(180,180)(-160,-10)
\put(0,0){\thicklines\line(1,0){170}} \put(0,150){\thicklines \line(1,0){170}}
\put(0,0){\thicklines\line(0,1){150}} \put(170,0){\thicklines \line(0,1){150}}
\put(25,10){\begin{picture}(20,20)
\multiput(00,110)(15,0){6}{\line(1,0){3}}
\put(0,110){\circle*{2}}
\put(-8,106){$1$}
\put(-8,20){$0$}
\put(24,126){$f(x)$}
\put(30,13){$x_{-}$}
\put(100,13){$x_{+}$}
\put(75,13){$x_{0}$}
\put(135,18){$x$}
\multiput(82,20)(0,15){6}{\line(0,1){3}}
\put(0,0){\vector(0,1){130}}
\put(0,20){\vector(1,0){130}}
\qbezier[1000](30,20)(100,200)(100,20)
\end{picture}}
\end{picture}
\caption{ Shape of a typical singularity spectrum.\label{singspec}}
\end{figure}
The characteristic feature of multifractal patterns in systems with disorder, like Anderson localisation transition and related phenomena, is the existence of essential sample-to-sample fluctuations of the prefactor $c_M(x)$ in different realizations of the disorder, as well as fluctuations in the number and height of extreme peaks of the pattern. Those fluctuations will be the subject of our interest.  At the same time the singularity spectrum $f(x)$ is typically a self-averaging convex function like the one shown in Fig.~2. Some general insight into statistical properties of disordered multifractals have been obtained in \cite{FLR2012} and the content of that work is concisely summarized below.

As is well-known \cite{CD,DL} disorder-generated multifractal patterns of intensities $h({\bf r})$ are typically {\it self-similar}, i.e.~characterized by the power-law correlation of intensities
\begin{equation}\label{5}
\fl \mathbb{E}\left\{h^q({\bf r}_1)h^s({\bf r}_2)\right\}\propto \left(\frac{L}{a}\right)^{y_{q,s}}\left(\frac{|{\bf r}_1-{\bf r}_2|}{a}\right)^{-z_{q,s}},\quad q,s\ge 0, \quad a\ll |{\bf r}_1-{\bf r}_2|\ll L,
\end{equation}
 and {\it spatially homogeneous}
\begin{equation}\label{6} \mathbb{E}\left\{h^q({\bf r})\right\}= \mathbb{E}\left\{\frac{1}{M}\sum_{\bf{r}} h^q({\bf r})\right\}\propto \left(\frac{L}{a}\right)^{d(\zeta_q-1)},
\end{equation}
where here and henceforth $\mathbb{E}\left\{A\right\}$ stands for the expected value (the mean) of the random variable $A$. The lattice model describes a situation where the relevant scales are $L$ and $a$, therefore it is natural to assume that intensities do not vary much over the scale $a$ and that they are uncorrelated at scale $L$. This can be expressed as
\begin{eqnarray}
\label{6b1}
\mathbb{E}\left\{h^q({\bf r}_1)h^s({\bf r}_2)\right\}&\sim \mathbb{E}\left\{h^{q+s}({\bf r}_1)\right\} &\qquad|{\bf r}_1-{\bf r}_2|\sim  a,\\
\label{6b2}
\mathbb{E}\left\{h^q({\bf r}_1)h^s({\bf r}_2)\right\}&\sim \mathbb{E}\left\{h^{q}({\bf r}_1)\right\}\mathbb{E}\left\{h^{s}({\bf r}_2)\right\}&\qquad |{\bf r}_1-{\bf r}_2|\sim  L.
\end{eqnarray}
If we make the further assumption that Eq.~\eqref{5} holds over the whole range $|{\bf r}_1-{\bf r}_2|\sim  a$ to $|{\bf r}_1-{\bf r}_2|\sim  L$, we directly get from \eqref{6b1}--\eqref{6b2} the relations between exponents
\begin{equation}\label{7}
 y_{q,s}=d(\zeta_{q+s}-1), \qquad z_{q,s}=d(\zeta_{q+s}-\zeta_{q}-\zeta_{s}+1),
\end{equation}
 so that the set of exponents $\zeta_q$ is the only one needed to characterize the spatial organization of such a multifractal pattern \cite{CD,DL}.

It proves to be instructive to shift the focus from the multifractal field $h({\bf r})$ to its logarithm $V({\bf r})=\ln{h({\bf r})}-\mathbb{E}\left\{\ln{h({\bf r})}\right\}$. Correlations of the field $V({\bf r})$ can be obtained by deriving $\langle h^qh^s\rangle-\langle h^q\rangle\langle h^s\rangle$, given by Eqs.~\eqref{5}--\eqref{6}, with respect to $q$ and $s$, using the identity $\frac{d}{ds}h^s|_{s=0}=\ln{h}$. Taking into account the relations \eqref{7} and the fact that $\zeta_0=1$ one arrives at the relation \cite{F10}
\begin{equation}
\label{8} \mathbb{E}\left\{V({\bf r_1})V({\bf r_2})\right\}=-d\,\zeta''_0\ln{\frac{|{\bf r}_1-{\bf r}_2|}{L}},
\end{equation}
where $\zeta''_0$ is the second derivative of $\zeta_q$ taken at $q=0$. We thus conclude that provided the conditions \eqref{5}--\eqref{6} of self-similarity and spatial homogeneity detailed above are fulfilled, the logarithm of a disorder-generated multifractal intensity must be necessarily a {\it log-correlated} random field \cite{F10}. Note that the nature of the higher cumulants is not fixed by this construction, and in particular there is no particular reason to expect Gaussianity of the field $V({\bf r})$ on general grounds. Moreover, had the field been Gaussian the only possible shape of the singularity spectrum $f(x)$ would be a simple parabola. In practice,  non-parabolic shapes are abundant in disordered multifractals \cite{EM2008}, although shapes extremely close to perfect parabolas also occur, most notably in the Integer Quantum Hall context \cite{QHparabolic}.

The shift of attention from the multifractal field to its logarithm is of conceptual and practical utility as extremes of random fields and processes with logarithmic correlations  attracted recently a lot of attention in physics \cite{CLD,FB,FLDR2009}, probability \cite{DZ,AZ}  and related areas. The most studied object is the 2D Gaussian free field (GFF) which is now believed to be as fundamental and rich as Brownian motion, and naturally emerges in studies ranging from quantum gravity and turbulence to financial mathematics. One of the most powerful rigorous frameworks for analyzing such fields and related processes relies upon the theory of "multiplicative chaos" \cite{VRrev}. Another important source of logarithmically correlated processes which frequently allow deep and rigorous analysis are   "hierarchic multiplicative cascades" originally suggested as a useful model of   turbulent velocity field \cite{Yaglom,Mandelbrot}. Later on, closely related models appeared in the context of polymers on disordered trees \cite{DS}. The latter model was realized to display a multifractal behaviour in \cite{CMW}. Yet independently and in a somewhat different version, cascades emerged as a model of multifractal eigenvectors of power-law banded matrices \cite{ME2000} based on the renormalization procedure suggested earlier in \cite{Levitov}.

\subsection{$1/f$ noises}
A few years ago it was realized that another representative of the same universality class are  1D processes known as $1/f$ noises.  Those (generalized) processes can be given {\it bona fide} mathematical definition as 1D "projections" of the 2D GFF or by various explicit constructions, for example as
a periodic random Gaussian process defined via the formal Fourier series
  \begin{equation} \label{9}
  V(t)=\sum_{n=1}^{\infty}
\frac{1}{\sqrt{n}} \left[v_n e^{i n t}+\overline{v}_n e^{-i n t}\right]\,,\quad t\in [0,2\pi),
\end{equation}
where  $v_n,\overline{v}_n$ are complex normal i.i.d.~variables with mean zero and variance unity.
As a simple calculation shows, the covariance structure of the process is logarithmic at small scale. More precisely,
 \begin{equation} \label{10}
\mathbb{E}\left\{V(t_1)V(t_2)\right\}= -2\ln|2\sin{\frac{t_1-t_2}{2}}|, \quad t_1\ne t_2.
\end{equation}
 One can also further consider aperiodic logarithmically-correlated processes such that
  $ \mathbb{E}\left\{V(t_1)V(t_2)\right\}\propto -\ln{|t_1-t_2|}$, as well as similar processes with stationary increments  with the structure function  $\mathbb{E}\left\{\left[V(t_1)-V(t_2)\right]^2\right\}\propto \ln{|t_1-t_2|}$ \cite{FKS}.  Among other things, such processes describe statistics of interesting mathematical objects: characteristic polynomials of random matrices and modulus of the  Riemann Zeta function along the critical line, on mesoscopic spectral scales \cite{FKS, FHK}.

\begin{figure}[h!]
\begin{center}
\includegraphics[width=0.5\textwidth,angle=0]{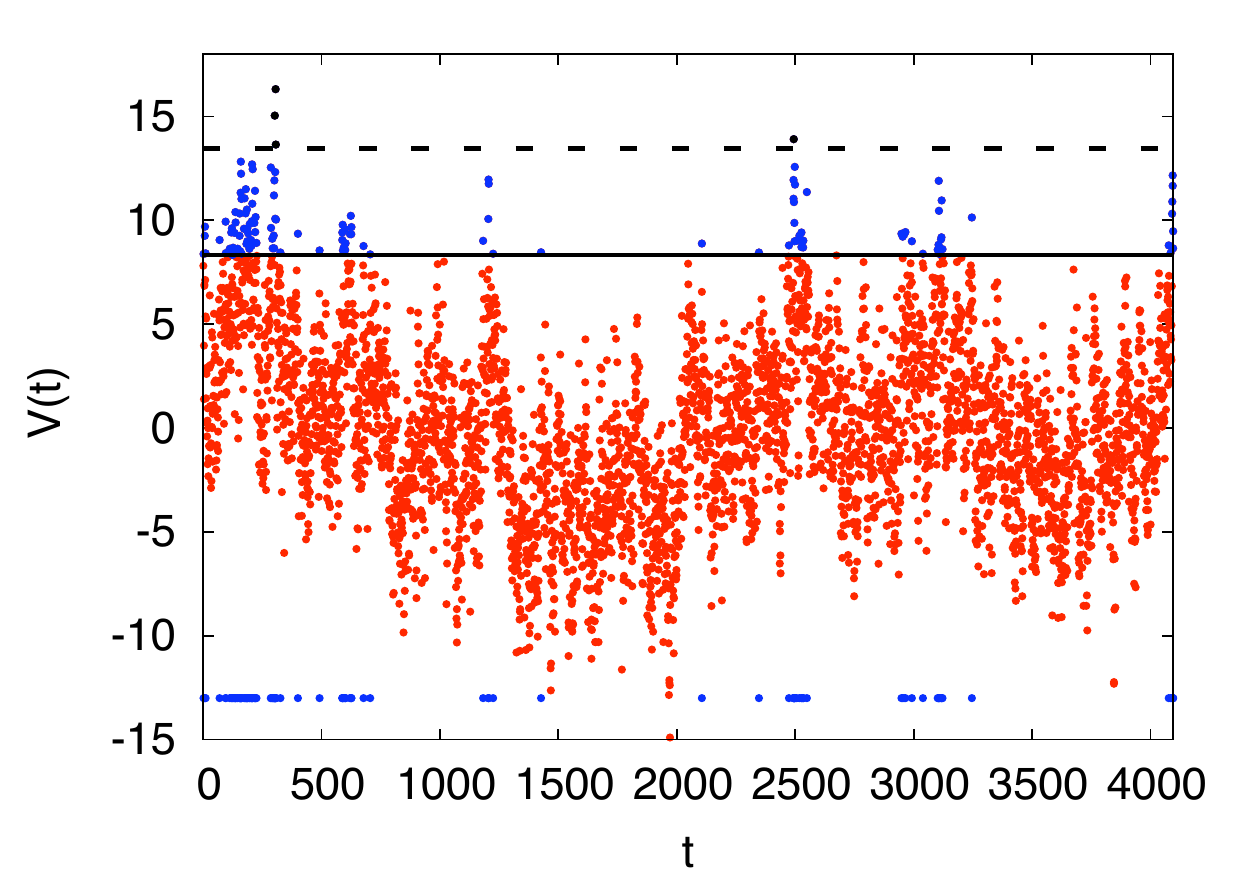}
\end{center}
 \caption{ The upper line marks the typical value of the {\it extreme value threshold} which for the present model is given by $V_m=2\ln{M}-\frac{3}{2}\ln{\ln{M}}$. The lower line is the  level $\frac{1}{\sqrt{2}}V_m$ and blue dots mark points supporting $V_i>\frac{1}{\sqrt{2}}V_m$ which form a manifestly fractal set. The figure is taken from \cite{FLR2012}.\label{extremethreshold}}
\end{figure}

  To understand statistics of high values and extremes of general logarithmically correlated random fields we will rely upon our intuition developed for the simplest 1D periodic case \eqref{9}. The process $V(t)$, rather than a random function of $t$, is a {\it random distribution}; therefore in practice it should be regularized.  There are several alternative regularizations. In particular, one can replace $V(t), \, t\in[0,2\pi)$ with a sequence of $M\gg 1$ random zero-mean Gaussian variables $V_k\equiv V\!\left(t=\frac{2\pi}{M}k\right)$ with a covariance matrix $C_{km}=\mathbb{E}\left\{V_k V_m\right\}$  given by
\begin{equation}\label{10a}
\fl \mathbb{E}\left\{V_k V_m\right\}=-2\ln{\left|2\sin{\frac{\pi(k-m)}{M}}\right|}, \quad C_{kk}=\mathbb{E}\left\{V_k^2\right\}>2\ln{M}, \quad \forall k=1,\ldots, M,
\end{equation}
where the inequality ensures that the matrix is positive definite. An example of a $1/f$ signal sequence generated for $M=4096$ for the discretized version of \eqref{9} is given at Fig.~\ref{extremethreshold}. The associated  multifractal intensity pattern is then generated by setting $h_i=e^{V_i}$ for each $ i=1,\ldots, M$.

Questions we would like to answer include: ({\bf i}) how many points are typically above a given level of the signal? ({\bf ii}) how strongly does this number fluctuate for $M\gg 1$ from one realization to the other? ({\bf iii}) How to understand the typical position $V_m$ and statistics of the {\it extreme values} (maxima or minima), etc. And, after all, what parts of the answers are expected to be  universal? The main advantage of the Gaussian $1/f$ noises is that it turns out to be possible to answer those questions in an explicit and detailed fashion \cite{FLR2012}, as follows.

When dealing with multifractal patterns it is frequently convenient to characterize them, as in \eqref{6}, by the set of exponents $\zeta_q$ describing the large-$M$ scaling behaviour of the so-called  partition functions
\begin{equation}\label{2}
Z_q=\sum_{i=1}^M\, h_i^q=\int_{-\infty}^{\infty}M^{q y}\rho_M(y)\,dy\sim M^{\zeta_q}\,, \quad \ln{M}\gg 1.
\end{equation}
Substituting for $\rho_M(y)$ the multifractal Ansatz (\ref{1}), the leading contribution to the integrals in the limit $\ln{M}\gg 1$ can be easily obtained by the Laplace method, yielding the asymptotic expressions for the counting function and for the partition function as
\begin{equation}\label{4}
{\cal N}_M(x)\approx \frac{c_M(x)}{|f'(x)|\sqrt{\ln{M}}}\,M^{f(x)}, \quad Z_q\approx \frac{c_M(y_*)}{\sqrt{|f''(y_*)|}} \, M^{\zeta_q},
\end{equation}
where $q$ and $y_*$ are related via the saddle-point condition $ f'(y_*)=-q$ and the exponents $\zeta_q$  are related to the singularity spectrum $f(x)$ by the  Legendre transform $\zeta_q=f(y_*)+q\,y_*$. The fluctuation properties of the counting function ${\cal N}_{M}(x)$ and the partition function $Z_q$ can therefore be related to each other via the statistics of the common prefactor $c_M(x)$. It turns out, as was discovered in \cite{FB,FLDR2009} (and independently from a different angle by Ostrovsky \cite{Ostrovsky}), that in the limit $\ln{M}\gg 1$ and for $|q|<1$, the positive integer moments $\mathbb{E}\left\{Z_q^n\right\}$ of the partition function $Z_q$ can be evaluated in a closed form in terms of the so-called  Selberg integrals \cite{FW}. It is then possible to derive the probability density of the random variable $Z_q$. For the particular case of the discrete periodic $1/f$ signal \eqref{10a} this distribution takes an especially simple form. As was shown in \cite{FB}, for $Z_{q}<M^2$ and $|q|<1$ one has
\begin{equation}\label{12}
 {\cal P}(Z_{q})=\frac{1}{q^2\, Z_e}\left(\frac{Z_e}{Z_q}\right)^{1+\frac{1}{q^2}}\,
e^{-\left(\frac{Z_e}{Z_{q}}\right)^{\frac{1}{q^2}}},\quad   Z_e=\frac{M^{1+q^2}}{\Gamma(1-q^2)}.
\end{equation}
The most important feature of this distribution is the forward power-law tail ${\cal P}(Z_{q})\sim Z_q^{-1-\frac{1}{q^2}}$ developed in a parametrically large region $Z_e\ll Z_q\ll M^2$. Defining the {\it typical} value ${\cal N}_t(x)$ of the counting function as $e^{\mathbb{E}\left\{\ln{\cal N}_M(x)\right\}}\sim {\cal N}_{t}(x)$, we introduce the scaled counting function $n={\cal N}_M(x)/{\cal N}_{t}(x)$, which measures the counting function against its characteristic scale. The counting function is correspondingly written in  the form ${\cal N}_M(x)=n \,{\cal N}_{t}(x)$, where ${\cal N}_{t}(x)$ is an averaged quantity and sample-to-sample fluctuations are now captured by the random variable $n$ (which depends on $x$ via its probability distribution). By exploiting Eq.~\eqref{4}, which relates the partition function $Z_q$ and the counting function ${\cal N}_M(x)$ via the function $c_M$, the probability density of $n$ can be obtained from (\ref{12}), giving \cite{FLR2012}
\begin{equation}\label{12a}
{\cal P}_x(n)=\frac{4}{x^2}\,
e^{-n^{-\frac{4}{x^2}}}\,n^{-\left(1+\frac{4}{x^2}\right)},\quad\,\quad 0<x<2
\end{equation}
for the distribution of the scaled counting function $n$, and
 \begin{equation}\label{12b}
{\cal N}_{t}(x)= \frac{M^{f(x)}}{x\sqrt{\pi\ln{M}}}\frac{1}{\Gamma(1-x^2/4)}, \qquad f(x)=1-x^2/4
 \end{equation}
for the typical value of the counting function. From \eqref{12a} one gets $\mathbb{E}\left\{ n \right\}=\Gamma(1-x^2/4)$, so that the characteristic scale ${\cal N}_t(x)$ is related to the {\it mean} value $\mathbb{E}\left\{{\cal N}_M(x)\right\}$ by
 \begin{equation}\label{12c}
{\cal N}_{t}(x)= \mathbb{E}\left\{{\cal N}_M(x)\right\}\frac{1}{\Gamma(1-x^2/4)}.
 \end{equation}
We see from (\ref{12c}) that for $x \to 2$ (that is for $x$ approaching the edge of the singularity spectrum support) the typical value ${\cal N}_{t}(x)$ is parametrically smaller than the mean value $\mathbb{E}\left\{{\cal N}_M(x)\right\}$ due to the diverging Gamma function factor in the denominator. By contrast, for {\it short-range} correlated random sequences the mean and the typical values of the counting function are always parametrically of the same order.

The position $x_m$ of the {\it typical} threshold of extreme values is determined from the natural condition ${\cal N}_{t}(x)\sim 1$. It is readily obtained from the expansion of \eqref{12b} in the vicinity of $x=2$; for logarithmically correlated processes it is given by
\begin{equation}\label{13}
x_m=2-c\,\frac{\ln{\ln{M}}}{\ln{M}}+O(1/\ln{M}) \quad \mbox{ with} \quad c=3/2.
\end{equation}
On the other hand, the value of $x$ determined by the condition $\mathbb{E}\left\{{\cal N}_M(x)\right\}\sim 1$ is given again by the formula \eqref{13} but with a different value $c=1/2$ which is a known universal value for short-range correlated sequences. The value $c=3/2$ was long conjectured to be a universal feature of systems with logarithmic correlations \cite{CLD}, and very recently there was a considerable progress of proving this fact with full mathematical rigour for a broad class of such systems \cite{BZ}. We believe that the above consideration of "typical vs.~mean" reveals a very transparent and intuitively clear mechanism behind such a universality \cite{FLR2012}. Such a difference is intimately connected to the existence of the power-law forward tail  $n^{-1-\frac{4}{x^2}}$ in the probability density (\ref{12a}), with the tail exponent approaching the value $-2$  when $x$ approaches the end of the support of the singularity spectrum $f(x)$. The tail is responsible for the diverging factor $\Gamma(1-x^2/4)\sim (2-x)^{-1}$  in \eqref{12c}, which eventually gives rise to the transmutation of $c=1/2$ into $c=3/2$.

\subsection{Disorder-generated multifractals}
The major features revealed in the above example are believed to be not specific for the Gaussian fields with logarithmic correlations  but can be further translated to generic disorder-generated multifractals. Indeed, though calculations of such generality are hardly feasible in the generic case, important insights into the statistical structure of such fields were obtained in the seminal paper of Mirlin and Evers \cite{ME2000}.

Namely, those authors considered a pattern of normalized multifractal weights $p_i=|\Psi_i|^2\sim M^{-\alpha_i}, \,\, i=1,\ldots M$, with $\Psi_i$ the wave-function components in models displaying multifractality due to the Anderson localisation transition phenomenon. To characterize such pattern they considered the moments (called  in that context  "inverse participation ratios" (IPR's))
\begin{equation}
\label{defipr}
I_q=\sum_{i=1}^Mp_i^q=\int_0^{\infty} M^{-q\alpha}\rho_M(\alpha)\,d\alpha,
\end{equation}
with $\rho_M(\alpha)$ the density of exponents $\alpha_i$. Note that IPR's are obvious analogues of the  partition functions $Z_q$, the only essential difference being the normalization condition $I_1=1$ and the (related)  condition of positivity of exponents $\alpha_i\ge 0$. The multifractality is reflected in the scaling exponent $\tau_q$ for the mean IPR via the scaling
\begin{equation}
\label{deftauq}
\mathbb{E}\left\{I_q\right\}\simeq \,M^{-\tau_q}.
\end{equation}

 Before going into detail related to \cite{ME2000} it is worth discussing another important feature of disorder-generated multifractals. Namely, one in general has to distinguish exponents $\tau_q$ from the set of {\it typical} exponents $\tauqtyp$ obtained from the scaling $\exp \mathbb{E}\left\{ \ln I_q \right\} \simeq M^{-\tauqtyp}$. We note that the exponents obtained from averaging the logarithm of either partition function or IPR are frequently called in the physical literature "quenched", whereas their counterpart extracted directly from averaged moments are known as "annealed". The existence of two different sets of exponents, $\tauqtyp$ (or quenched) versus
$\tau_q$ (or annealed), governing correspondingly the scaling behaviour of
typical $I_q$ versus disorder averaged IPR's is the generic feature of multifractality in the presence of disorder.
The possibility of "annealed" average to produce results different from typical is related
to a possibility of disorder-averaged moments to be dominated by
exponentially rare configurations in some parameter range.  The singularity spectrum $f(\alpha)$ is related to the multifractal exponents $\tau_q$ by the Legendre transform $f(\alpha)=\min_q(q\alpha -\tau_q)$. The typical singularity spectrum $\ftyp(\alpha)$ always has terminating points at $\ftyp(\alpha_{\pm})=0$ and is supported by the interval $[\alpha_{-}, \alpha_{+}]$, see Fig.~2. In contrast to the typical singularity spectrum
the "annealed" version of the spectrum recovered from the exponents $\tau_q$ via the Legendre transform extends beyond the support interval $[\alpha_{-}, \alpha_{+}]$ and is negative there: $f(\alpha)<0$, see Fig.~4.
\begin{figure}[h]
\begin{picture}(200,200)(-160,-30)
\put(0,-20){\thicklines\line(1,0){170}} \put(0,150){\thicklines \line(1,0){170}}
\put(0,-20){\thicklines\line(0,1){170}} \put(170,-20){\thicklines \line(0,1){170}}
\put(25,10){\begin{picture}(20,20)
\multiput(00,110)(15,0){6}{\line(1,0){3}}
\put(0,110){\circle*{2}}
\put(-8,106){$1$}
\put(-8,20){$0$}
\put(4,126){$f(\alpha)$}
\put(30,13){$\alpha_{-}$}
\put(100,13){$\alpha_{+}$}
\put(75,13){$\alpha_{0}$}
\put(135,18){$\alpha$}
\multiput(82,20)(0,15){6}{\line(0,1){3}}
\qbezier[1000](30,20)(100,200)(100,20)
\multiput(14,-19)(1,2.5){16}{\circle*{1}}
\multiput(100,20)(0.15,-4.5){8}{\circle*{1}}
\put(0,-10){\vector(0,1){140}}
\put(0,20){\vector(1,0){130}}
\end{picture}}
\end{picture}
\caption{Shape of an "annealed" multifractality spectrum with negative parts (dotted) extracted from the disorder-averaged moments
and reflecting exponentially rare events, see the text.\label{singspecannealed}}
\end{figure}
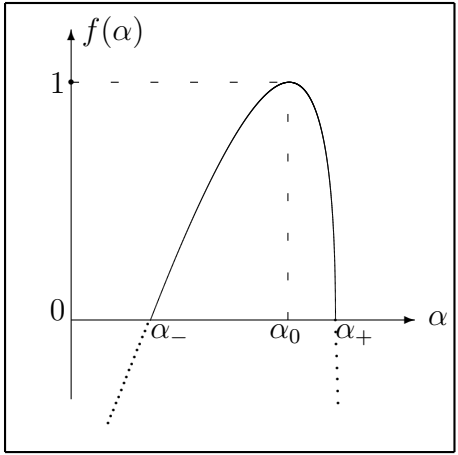
Indeed, those values reflect events which are exponentially rare \cite{negf} and
need exponentially many realisations of disorder to be observed
experimentally or numerically. On the other hand, when dealing with {\it typical} multifractality spectrum $f^{typ}(\alpha)$ by exploiting the relation (\ref{defipr}) it seems natural to specify the limits of integration over $\alpha$ to be precisely $\alpha_{-}\le \alpha \le \alpha_+$. Typical IPR moments are then given by
\begin{equation}\label{typ1}
I^{typ}_q=\int_{\alpha_{-}}^{\alpha_{+}}\, M^{-q\alpha+f^{typ}(\alpha)} d\alpha \sim M^{-\tau^{typ}_q}\,,
 \end{equation}
 and calculating the above integral by the steepest descent method reveals that typical (or {\it quenched})
 exponents $\tau_q^{typ}$
 are related to $f^{typ}(\alpha)$ by Legendre transform only in the range
 $\frac{df}{d\alpha}|_{\alpha_{+}}=q_{min}\le q\le q_{max} =\frac{df}{d\alpha}|_{\alpha_{-}}$, whereas outside that interval the integral is dominated by the boundaries and the exponents must behave linearly in $q$, that is $\tau^{typ}_q=q\alpha_{\pm}$, see Fig.~5.

 \begin{figure}[h]
\begin{picture}(150,170)(-120,-10)

\put(0,0){\thicklines\line(1,0){150}} \put(0,150){\thicklines \line(1,0){150}}
\put(0,0){\thicklines\line(0,1){150}} \put(150,0){\thicklines \line(0,1){150}}
\put(50,5){\vector(0,1){130}}
\put(5,70){\vector(1,0){130}}

\multiput(105,80)(4,0.7){7}{\circle*{1}}
\multiput(27,5.5)(2,3.5){5}{\circle*{1}}

\put(25,10){\begin{picture}(20,20) \put(10,59){\circle*{1.5}}
\put(80,59){\circle*{1.5}} \put(22,126){$\tau^{typ}_q$}
\put(4,52){$q_{min}$} \put(6,28){$-1$} \put(55,52){$1$}
\put(80,52){$q_{max}$} \put(114,58){$q$}
\qbezier[2000](10,10)(40,60)(80,70)
\end{picture}}
\end{picture}
\caption{$q$-dependence of typical ("quenched") multifractality exponents $\tau_q$. Dotted lines show linear behaviour, see the text.}
\end{figure}
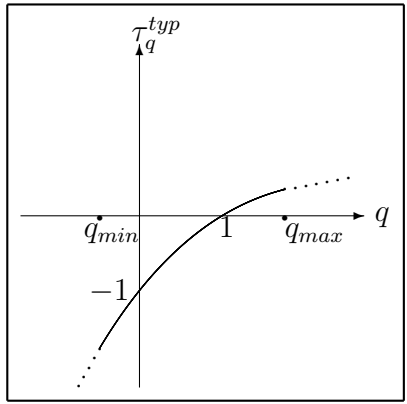

We will not dwell on further interesting differences of "quenched" vs. "annealed"
exponents and direct the interested reader to \cite{F09} for some related discussion and references.

Returning to  Mirlin and Evers paper \cite{ME2000}, those authors suggested that the probability density of IPR's should have the scaling form ${\cal P}_q(n)$,  with the scaling variable $n=I_q/I_q^{(t)}$ set by the characteristic scale $I_q^{(t)}$ which is simultaneously the  {\it typical} IPR value. Moreover, they argued that the distribution ${\cal P}_q(n)$  should display a power-law tail ${\cal P}_q(n)\sim n^{-1-\omega_q}, \,\,n\gg 1$. Assuming this picture let us denote $\overline{n}_q=\int_0^{\infty}{\cal P}_q(n)\, n\, dn$. Note that now we can rewrite (\ref{deftauq}) conveniently as $\mathbb{E}\left\{I_q\right\}= I_q^{(t)} \overline{n}_q\simeq B(q)\,M^{-\tau_q}$, with $B(q)$ some coefficient. For any function $\phi_q$ of the variable $q$ we can further define the "conjugate" function $\phi_*(\alpha)$ by the relation $\phi_*\left(\alpha(q)\right)=\phi_q$. We can then cast the multifractal Ansatz for the density of exponents in the same form as \eqref{1}, namely
\begin{equation}\label{14}
\rho_M(\alpha)=\sum_{i=1}^M\,\delta\left(\frac{\ln{p_i}}{\ln{M}}-\alpha\right)\approx \frac{n_*(\alpha)
}{\overline{n}_*(\alpha)}B_{*}(\alpha)\sqrt{\frac{\ln{M}|f''(\alpha)|}{2\pi}}\, M^{f(\alpha)},
\end{equation}
where $n_*(\alpha)$ is a random coefficient of the order of unity distributed according to the probability density  ${\cal P}^*_{\alpha}(n)$ defined via the rule ${\cal P}^*_{\alpha(q)}(n)={\cal P}_q(n)$. In order to show \eqref{14}, one can simply perform the integral in \eqref{defipr} by the Laplace method and check that it gives back precisely the value $I_q=I_q^{(t)}n_q$, with $n_q=n_*(\alpha(q))$.

 Now we can substitute the Ansatz (\ref{14}) to the definition of the counting function $N_<(\alpha)=\int_{-\infty}^{\alpha}\rho_M(\alpha)\,d\alpha$, choosing $\alpha$ to the left of the maximum of $f(\alpha)$ (that is, $\alpha_{-}<\alpha<\alpha_0$), and perform the integral by the Laplace method. As a result, similarly as in the previous section, the counting function can be put asymptotically under the form ${\cal N}_{<}(\alpha)\simeq n_*(\alpha) {\cal N}_t(\alpha)$, where the scale ${\cal N}_t(\alpha)$ defines the typical value of the counting function and is given by
\begin{equation}\label{15}
{\cal N}_t(\alpha)=\frac{B_{*}(\alpha)}{\overline{n}_*(\alpha)f'(\alpha)}\sqrt{\frac{|f''(\alpha)|}{2\pi\ln{M}}}\, M^{f(\alpha)}, \quad \alpha_{-}<\alpha<\alpha_0.
\end{equation}
 The typical maximal value among $p_i$'s in the multifractal pattern is then given by $p_m=M^{-\alpha_m}$, where $\alpha_m$ is determined from the condition that the typical value of the counting function becomes of the order of unity, that is, ${\cal N}_t(\alpha_m)\sim 1$. Similarly as in the previous section, the position $\alpha_m$ of the typical threshold will be obtained by expanding \eqref{15} in the vicinity of $\alpha=\alpha_{-}$, corresponding to $q\to q_{c}\equiv f'(\alpha_-)$ (which is now the left termination point of the singularity spectrum rather than the right one in the previous section, given the sign differences in the definition of exponents $\zeta_q$ and $\tau_q$). Generically at the edge we expect that $f'(\alpha_{-}), |f''(\alpha_{-})|$ and $B_{*}(\alpha_{-})$ be all finite and positive.  On the other hand, Mirlin and Evers argued that the tail exponent $\omega_{q}$ featuring in the probability density ${\cal P}_q(n)\sim n^{-1-\omega_q}$  must tend to the value $\omega_{q_{c}}=1$ when $q\to q_{c}$. This immediately implies that there must be a divergence of the mean value $\overline{n}_q=\int_0^{\infty}{\cal P}_q(n)\, n\, dn$ for $q\to q_c$, generically as $\overline{n}_{q\to q_c}\sim |q-q_c|^{-1}$; in turn, this divergence is translated into the threshold behaviour $\overline{n}_*(\alpha)\sim (\alpha-\alpha_{-})^{-1}$. The analog of Eq.~\eqref{12c} is $\mathbb{E}\{{\cal N}_{<}(\alpha)\}=\overline{n}_*(\alpha) {\cal N}_t(\alpha)$. The term $\overline{n}_*(\alpha)$ plays a similar role as the $\Gamma(1-x^2/4)$ term in the previous section and is responsible for the appearance of a factor $3/2$ in the extreme value statistics. Indeed, approximating $f(\alpha_m)\approx f'(\alpha_{-})(\alpha_m-\alpha_{-})$  we immediately find from \eqref{15} and the condition ${\cal N}_t(\alpha_m)\sim 1$ that  the threshold $\alpha_m$ must be given to the first non-trivial order by
 \begin{equation}\label{16}
\fl \alpha_m\approx \alpha_{-}+\frac{3}{2}\frac{1}{f'(\alpha_{-})}\frac{\ln{\ln{M}}}{\ln{M}} \quad \Rightarrow -\ln{p_m}\approx \alpha_{-}\ln{M}+ \frac{3}{2}\frac{1}{f'(\alpha_{-})}\ln{\ln{M}},
 \end{equation}
which is the analog of \eqref{13}. For branching random walks this result has been indeed rigorously proved recently \cite{A-BR,Aidekon}.

\subsection{Behaviour for $|q|>1$}
 Finally, we very briefly discuss the behaviour related to statistics of the {\it extreme} (i.e.~largest/smallest) values of logarithmically correlated sequences and processes \cite{CLD,FB,FLDR2009,DZ,BZ}. If we come back to the partition function $Z_q$ of the Gaussian $1/f$ model, extreme values start to dominate its distribution in the region $q>1$, so that the probability density of $Z_q$ changes qualitatively. Defining the scaled moments for $q>1$ as $z=Z_{q}\,\left(\frac{(\ln{M})^{3/2}}{M^2}\right)^{q}$, the most salient feature of the density ${\cal P}_q(z)$  for $\ln{M}\gg 1$ is predicted to be the following tail \cite{CLD}:
  \be \label{momdisslow}
 {\cal P}_q(z)\propto z^{-\left(1+\frac{1}{q}\right)}  \ln{z}, \quad z\gg 1, \quad q>1.
\ee
Both the change of the tail exponent from $1+\frac{1}{q^2}$ to $1+\frac{1}{q}$ and the presence of the logarithmic factor $\ln{z}$
in (\ref{momdisslow}) are different manifestations of the so-called {\it freezing transition} occurring at $q=1$
and believed to be a universal feature of generic random processes with logarithmic correlations \cite{F10,CLD,FB,ZeitouniFreezing}. It would be therefore also consistent to expect a similar behaviour of the IPR's, that is, ${\cal P}(I_q)\propto I_q^{-\left(1+\frac{q_c}{q}\right)}\,\ln{I_q}$ for $q>q_c$.

The knowledge of the distribution of the partition function for $q\to \infty$ is equivalent to the knowledge of the
probability density of the highest maximum $V_m$ of the field. To the leading and subleading order the position of the highest maximum is clearly the same as the "extreme threshold level", see (\ref{13}) and Fig.~\ref{extremethreshold}, so we can conveniently parametrize
\begin{equation}
V_m=2\ln{M}-\frac{3}{2}\ln{\ln{M}}+y,
\end{equation}
where $y$ is a random variable of order of unity. In general the probability density of $y$ is not known explicitly in full detail, but for the simplest model (\ref{10a}) it is conjectured to have the form:
\begin{equation}\label{prediction1}
p(y)=-\frac{d}{dy}\left[2 e^{y/2} K_1(2 e^{y/2})\right]=2e^{y}K_0(2 e^{y/2}),
\end{equation}
where $K_{\nu}(z)$ denotes the modified Bessel function of the second kind. The asymptotic behaviour $p(y\to -\infty)\approx -y e^y+\ldots$ is conjectured to be the
universal backward tail shared (after appropriate rescaling)  by extreme value distribution of generic logarithmically-correlated random processes, see \cite{CLD}. This is manifestly different from the short-ranged random processes characterized by Gumbel distribution of extremes with the corresponding backward tail $p(y\to -\infty)\approx e^y$.

\section{Ruijsenaars-Schneider model}
We now illustrate the above points on an example.
\begin{figure}[t]
\begin{center}
\includegraphics*[width=.65\linewidth]{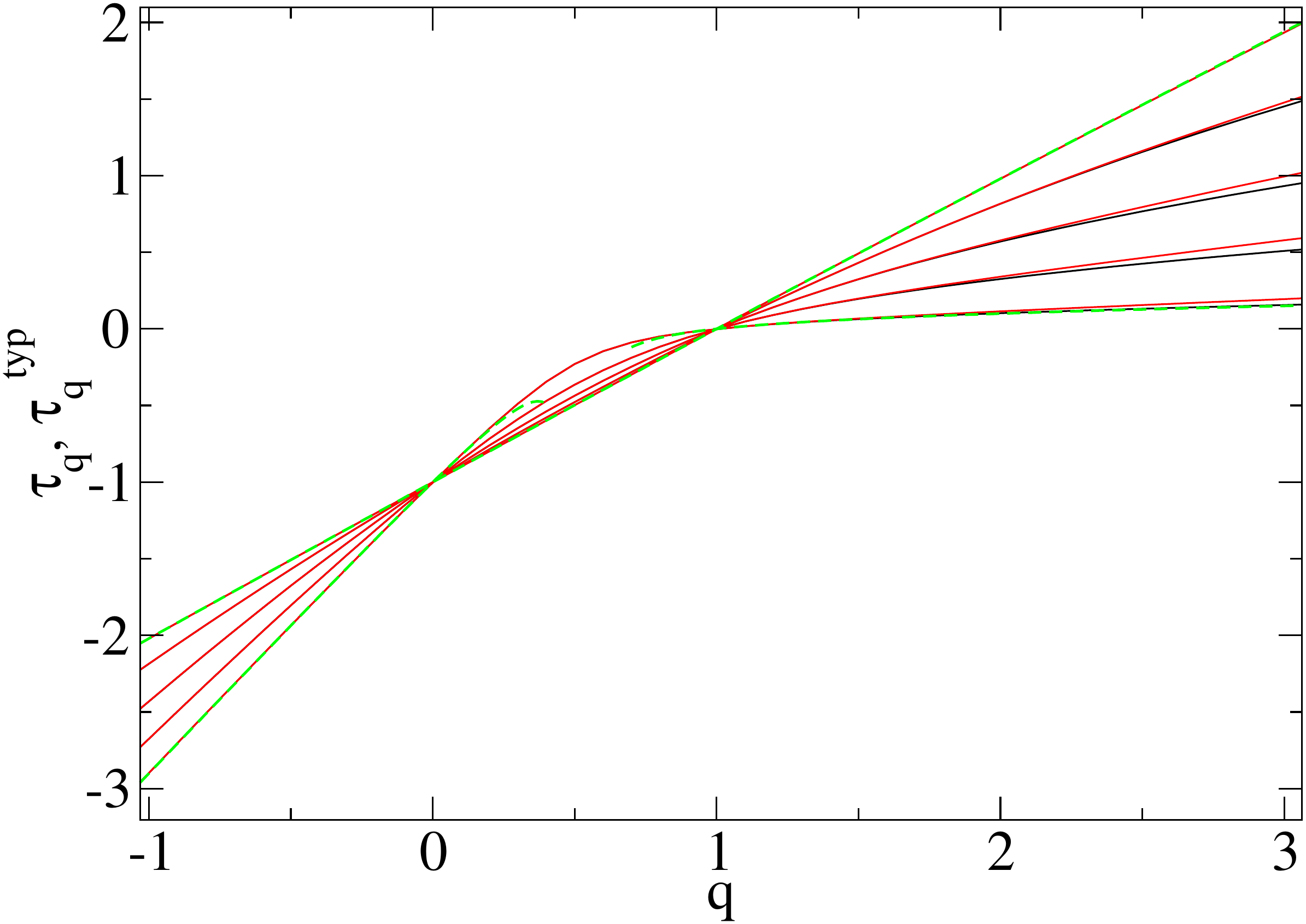}
\end{center}
\caption{Multifractal exponents $\tau_q$ (black) and $\tauqtyp$ (red/grey) for the random matrix ensemble \eqref{laxruij}, from bottom to top on the right $a=0.1, 0.3, 0.5, 0.7, 0.9$, extracted from matrices of size $M=2^{n}$ with $n$ ranging from $7$ to $12$, and averaged over $2^{18-n}$ matrix realizations. Dashed green lines are the analytic expressions \eqref{dqstrong} and \eqref{dqweak}.
\label{figdq_ruij}}
\end{figure}
The Ruijsenaars-Schneider model \cite{ruijsc} is an integrable model describing the motion of $M$ classical relativistic particles on a line which generalizes the nonrelativistic Calogero-Moser models. The classical dynamics is characterized by the Hamiltonian
\begin{equation}
H({\bf p},{\bf q})=\sum_j\cos(p_j)\prod_{k\neq j}\left (1-\frac{\sin^2\tau}{\sin^2 [\frac{q_j-q_k}{2}]}\right)^{\frac12},
\end{equation}
where $q_i$ and $p_i$, $1\leq i\leq M$, are positions and momenta of the particles, and $\tau$ is some parameter. One can show that the equations of motion are equivalent to $\dot{L}=T\,L-L\,T$, where $L,T$ is a pair of Lax matrices of size $M\times M$, and the dot denotes time derivative. In \cite{BogGirPRL}, an ensemble of unitary random matrices was constructed from the Lax matrix $L$, namely the ensemble of matrices
\begin{equation}
\label{laxruij}
L_{jk}=\frac{e^{i \Phi_j}}{M}\frac{1-e^{2i\pi a }}{1-e^{2i\pi(j-k+a)/M}},
\end{equation}
with $\Phi_j$ independent random variables uniformly distributed in $[0,2\pi]$ and $a$ a real parameter. The model is integrable, and explicit action-angle transformations can be obtained \cite{rui95}, which allows to calculate the joint probability distribution for the eigenvalues of $L$ \cite{BogGirSch11}. Spectral statistics turn out to be of intermediate type, with level repulsion at small spacings and exponential decay.
\begin{figure}[t]
\begin{center}
\includegraphics*[width=.65\linewidth]{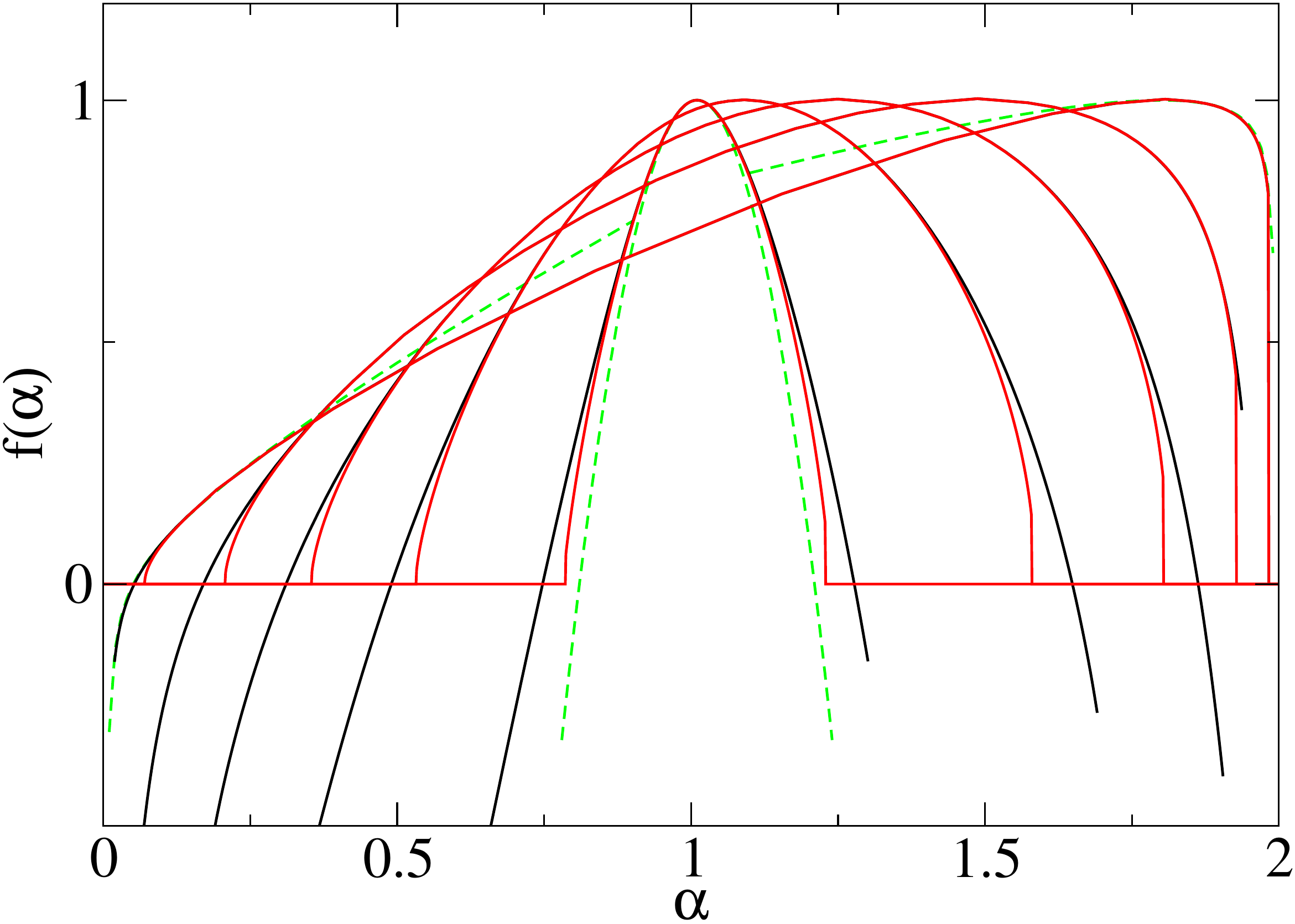}
\end{center}
\caption{Singularity spectrum $f(\alpha)$ (black) and $\ftyp(\alpha)$ (red/grey) for the ensemble \eqref{laxruij}, from widest to narrowest $a=0.1, 0.3, 0.5, 0.7, 0.9$, obtained from Legendre transform of $\tau_q$, data from Fig.~\ref{figdq_ruij}. Dashed green lines are the Legendre transform of the analytic expressions \eqref{dqstrong} and \eqref{dqweak}.\label{figfq_ruij}}
\end{figure}
Multifractality of eigenvectors of \eqref{laxruij} has been investigated in \cite{BogGir11, BogGir12}.  Analytic expressions for multifractal exponents $\tau_q$ defined by \eqref{deftauq} have been obtained for values of the parameter $a$ close to an integer. When $a$ is close to zero the lowest-order terms give
\begin{eqnarray}
\label{dqstrong}
&\tau_q=\frac{2a}{\sqrt{\pi}}\frac{\Gamma(q-\frac12)}{\Gamma(q-1)}\hfill&q>\frac12\nonumber\\
&\tau_q=2q-1-\frac{2a\,q}{\sqrt{\pi}}\frac{\Gamma(\frac12-q)}{\Gamma(1-q)}\;\;\;\;\;\;&q<\frac12,
\end{eqnarray}
while for $a$ close to a nonzero integer $k$ one finds
\begin{equation}
\label{dqweak}
\tau_q=q-1-q(q-1)\frac{(a-k)^2}{k^2},
\end{equation}
which corresponds to a parabolic singularity spectrum.
\begin{figure}[t]
\begin{center}
\includegraphics*[width=.48\linewidth]{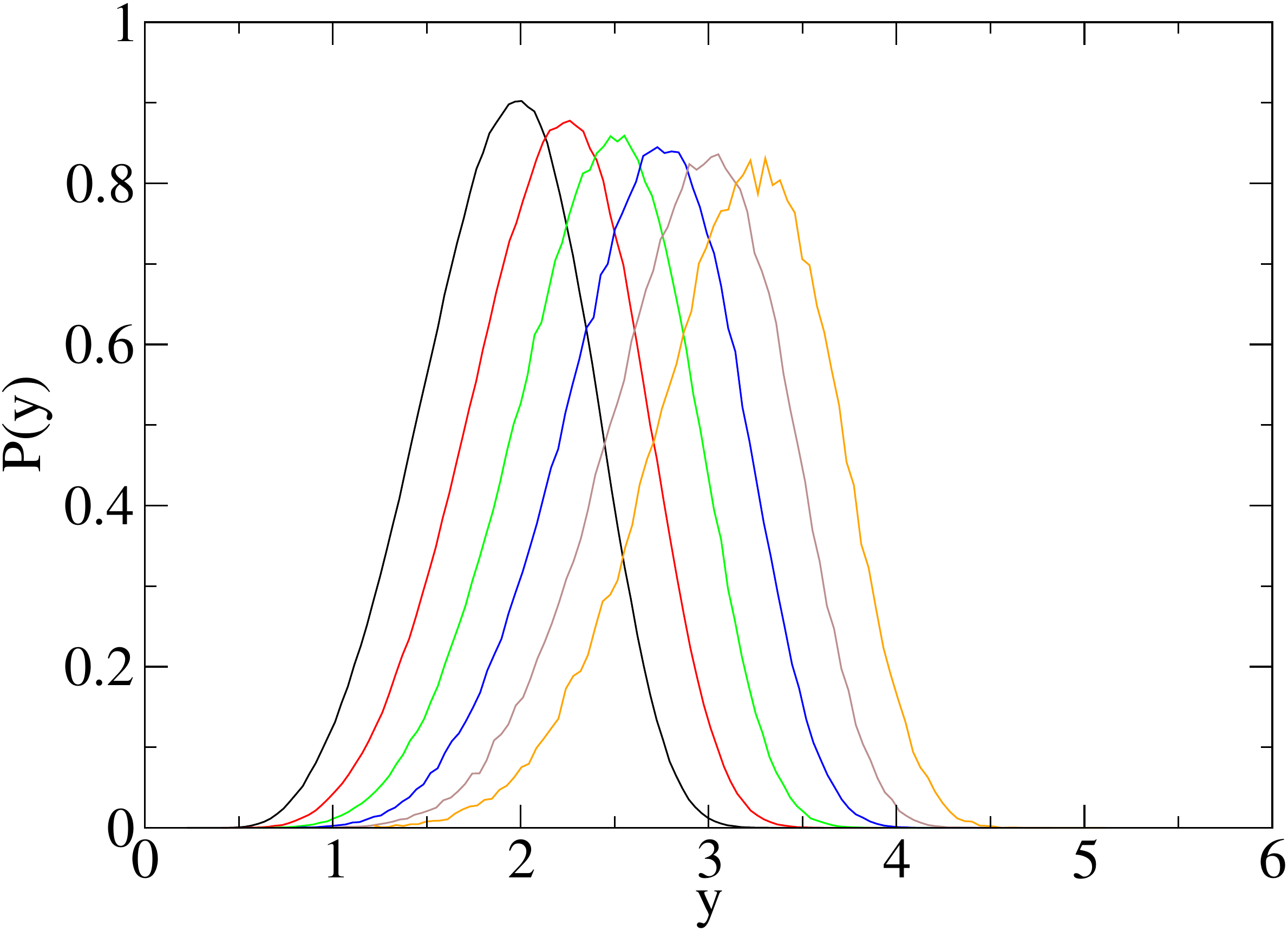}
\includegraphics*[width=.48\linewidth]{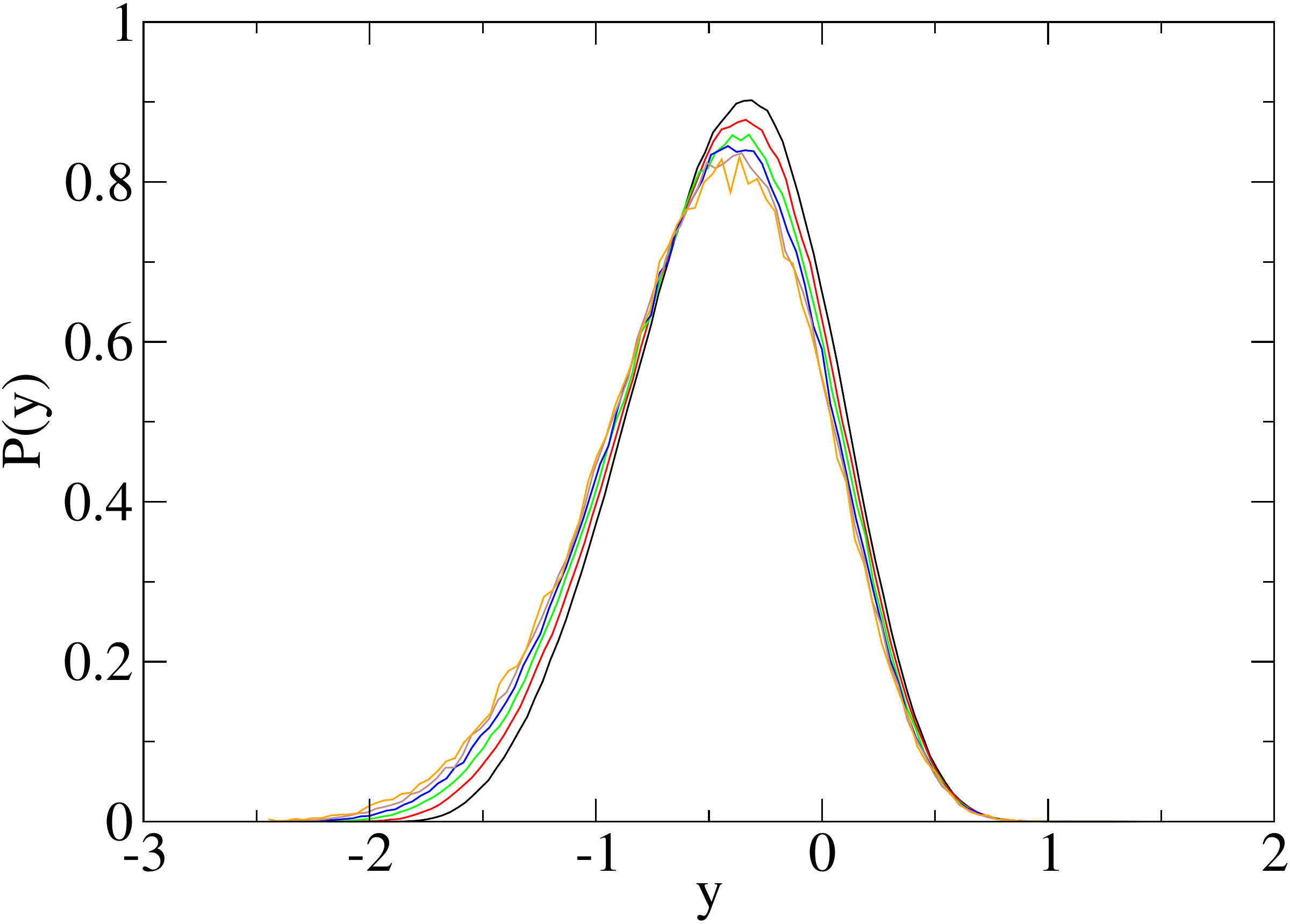}\\
\includegraphics*[width=.48\linewidth]{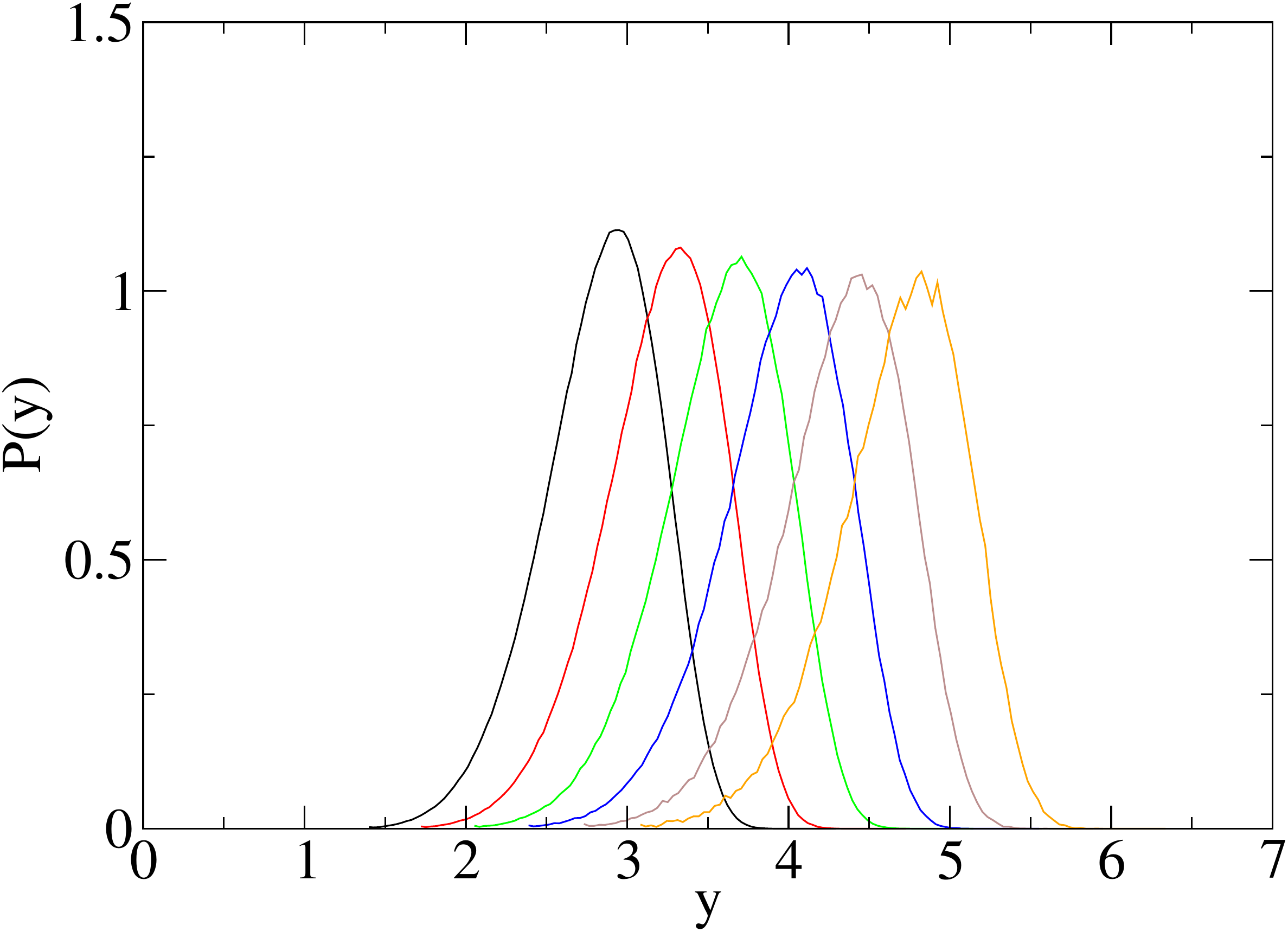}
\includegraphics*[width=.48\linewidth]{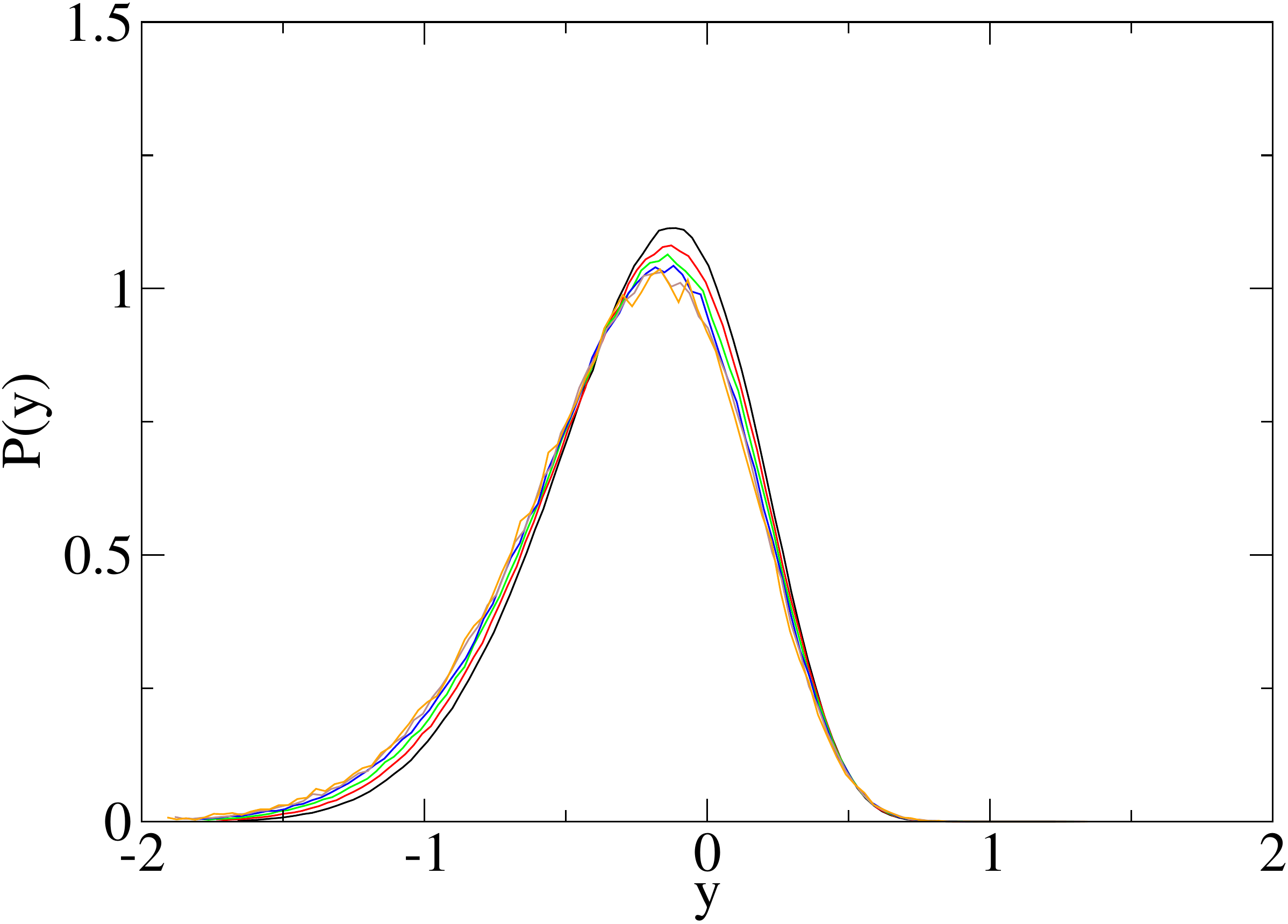}\\
\includegraphics*[width=.48\linewidth]{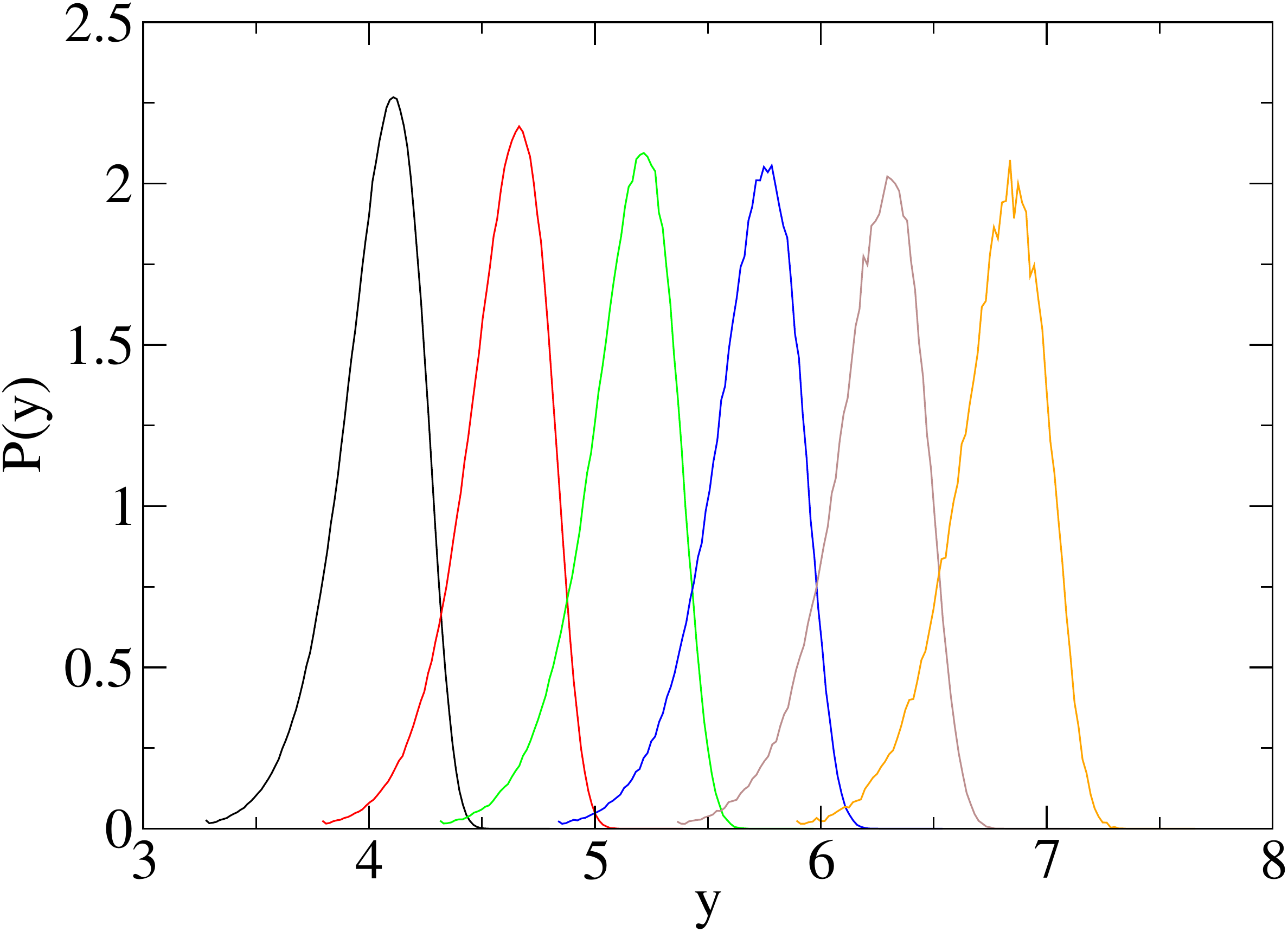}
\includegraphics*[width=.48\linewidth]{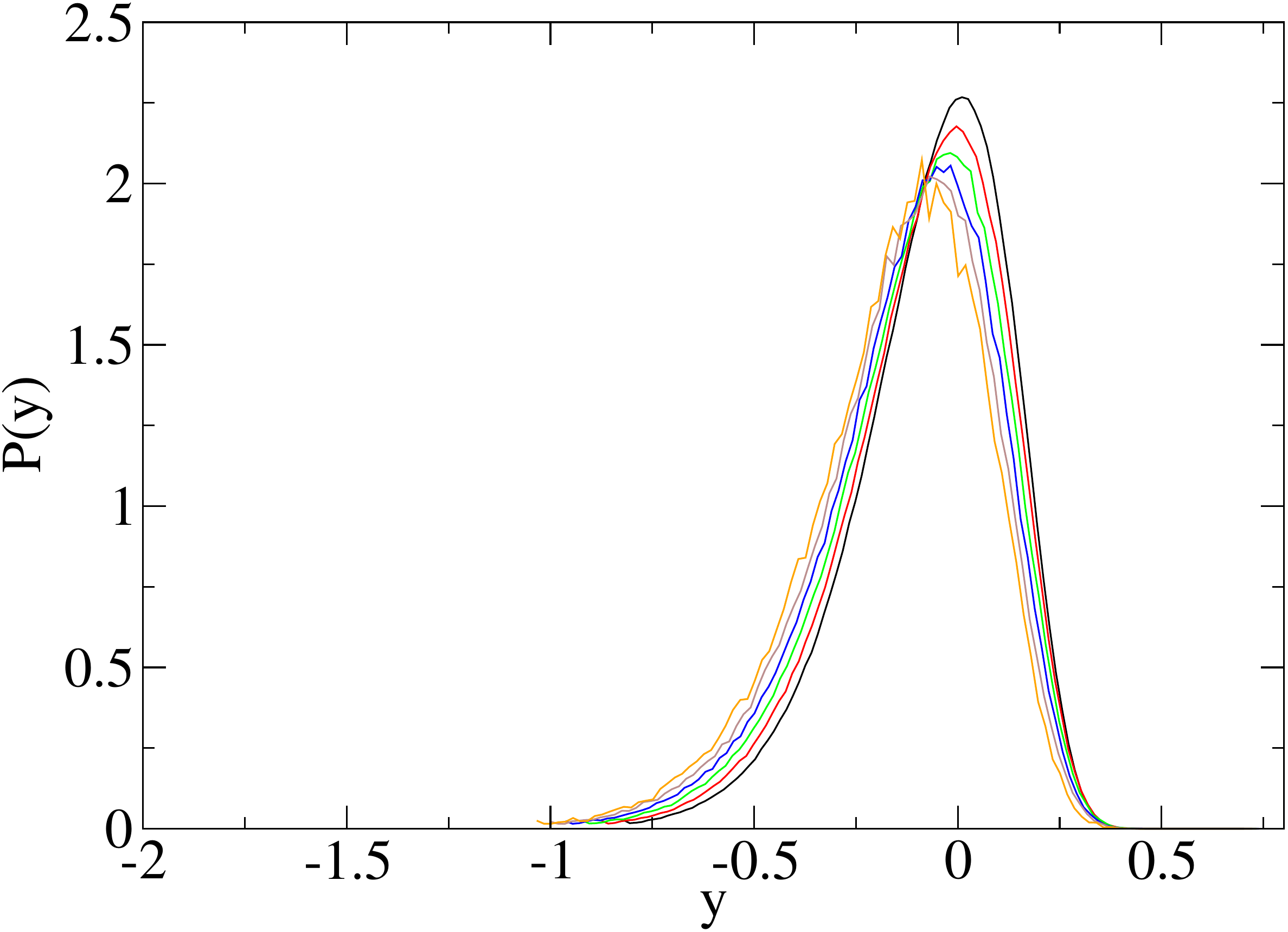}\\
\end{center}
\caption{Distribution of $y=-\ln p_m$ for eigenvectors of matrices from the ensemble \eqref{laxruij} for $a=0.5$ (top), $0.7$ (middle) and $0.9$ (bottom), and $M=2^{n}$ with $n=7$ to $12$ (from black to orange). Data from $2^{15}$ down to $24$ realizations. Left: original data. Right: shift $y\to y-\alpha_{-}\ln M-\frac{3}{2f'(\alpha_{-})}\ln\ln M$, with values of $\alpha_{-}$ and $f'(\alpha_{-})$ extracted from $f(\alpha)$ in Fig.~\ref{figfq_ruij}. \label{pdemax_ruij}}
\end{figure}
To go beyond these two regimes one resorts to numerical simulations. Numerical results are shown for the exponents $\tau_q$ and $\tauqtyp$ in Fig.~\ref{figdq_ruij}, and for the corresponding singularity spectrum in Fig.~\ref{figfq_ruij} for various values of $a$. For $a$ close to 0 (almost localized states), the plot obtained from the analytic expression \eqref{dqstrong} coincides with the numerical singularity spectrum $f(\alpha)$. On the other hand, in the weak multifractality limit $a\simeq 1$ the singularity spectrum cannot be extracted reliably in the region corresponding to large $q$, as can be seen in Fig.~\ref{figfq_ruij}, where the numerical curves significantly depart from the analytic expression in this regime. This comes from the fact that states are nearly extended, so that vector components are of order $\sim 1/M$; for large values of $q$ (corresponding to the bottom of the plot), moments become tiny and numerical errors become significant, which yields a highly fluctuating $\tau_q$ and makes it difficult to identify a minimum in the Legendre transform.\\

In the case of a generic disorder-generated multifractal pattern, for each eigenvector $\Psi$ the quantity $y=-\ln p_m$, with $p_m=\max_{1\leq i \leq M}|\Psi_i|^2$, is expected to follow Eq.~\eqref{16}. The values for $\alpha_{-}$ and $f'(\alpha_{-})$ can be extracted either from the intersection points of $f(\alpha)$ with the $x$--axis or from the termination points of the typical singularity spectrum $\ftyp(\alpha)$. The results for the distribution of $y$, with values extracted from the plot of Fig.~\ref{figfq_ruij} in the parameter region where they can be obtained more reliably, are shown in Fig.~\ref{pdemax_ruij}, showing the accuracy of Eq.~\eqref{16} for our model.

\section{Correlations in the Ruijsenaars-Schneider model}
 The goal of this section is to further substantiate the idea of a close relationship between the logarithmically-correlated fields and multifractality by revealing the hidden logarithmic structure of the RS model. To this end we explicitly perform {\it ab initio} evaluation of the two-point correlation function (i.e. covariance) of the logarithm of the multifractal intensity, that is $\mathbb{E}\left\{ V_iV_j\right\}$, with $V_i=\ln|\Psi_i|^2-\mathbb{E}\left\{\ln|\Psi_i|^2\right\}$. To make the calculation analytically tractable we consider perturbation expansion of \eqref{laxruij} around an integer $\kappa$, setting $a=\kappa+\epsilon$ with small expansion parameter $\epsilon\ll 1$. We set $U_{mn}=L_{mn}e^{-i\pi \epsilon (1-1/M)}$. Doing this results in rescaling $L_{mn}$ by a trivial factor, but has the advantage that the matrix $U$ can now be expressed as
\begin{equation}
\label{rescale}
U_{mn}=\delta_{m-n+\kappa}\frac{e^{i\Phi_m}}{M}\frac{\sin\pi \epsilon}{\sin(\pi \epsilon/M)}
+(1-\delta_{m-n+\kappa})\frac{e^{i\Phi_m}}{M}\frac{(1-e^{2\pi i \epsilon })e^{-i\pi \epsilon (1-1/M)}}{1-e^{2\pi i(m-n+\kappa+\epsilon)/M}},
\end{equation}
so that both terms in the above expression have a definite limit when $\epsilon\to 0$. First-order expansion of $U_{mn}$ reads
\begin{equation}
\label{M1RS}
U_{mn}\simeq e^{i\Phi_m}\delta_{m-n+\kappa}-\frac{2i\pi\epsilon}{M}e^{i\Phi_m}\frac{1-\delta_{m-n+\kappa}}{1-e^{2\pi i(m-n+\kappa)/M}}\ .
\end{equation}
For simplicity we consider the case $\kappa=1$. Eigenstates of $U$ are labeled by $\alpha$, $1\leq \alpha\leq M$. Unperturbed eigenstates, that is, eigenvectors of $e^{i\Phi_m}\delta_{m-n+1}$, are given by
\begin{equation}
\Psi^{(0)}_n(\alpha)=\frac{1}{\sqrt{M}}e^{i S_n(\alpha)},\qquad
 S_n(\alpha)=\frac{2\pi}{M}n\alpha+n\tilde{\Phi}-\sum_{j=0}^{n-1}\Phi_{j}
\label{psi0}
\end{equation}
with eigenvalues
\begin{equation}
\lambda_{\alpha}^{(0)}=e^{i \tilde{\Phi}+\frac{2i\pi}{M}\alpha} \quad \mbox{where} \quad \tilde{\Phi}=\frac1M\sum_{j=0}^{M-1}\Phi_j\,.
\end{equation}
The  first-order perturbation expansion gives
\begin{equation}
\Psi_n(\alpha)=\Psi^{(0)}_n(\alpha)+\sum_{\beta}C_{\alpha \beta}\Psi^{(0)}_n(\beta),
\label{Psi_n_alpha}
\end{equation}
with
\begin{equation}
\label{cabb}
C_{\alpha \beta}=\frac{\langle\Psi^{(0)}(\beta)|U^{(1)}|\Psi^{(0)}(\alpha)\rangle}{\lambda_{\alpha}^{(0)}-\lambda_{\beta}^{(0)}}
\end{equation}
and $U^{(1)}$ denotes the order-$\epsilon$ (off-diagonal) term in \eqref{M1RS}. Replacing $\Psi^{(0)}$ by its explicit value \eqref{psi0} we get from \eqref{Psi_n_alpha}
\begin{equation}
|\Psi_n(\alpha)|^2=\frac{1}{M}\left(1+Q_n(\alpha)+Q_n^{*}(\alpha)+Q_n(\alpha)Q_n^{*}(\alpha)\right)
\label{psi2bis}
\end{equation}
with
\begin{equation}
\label{defqn}
Q_n(\alpha)=\sum_{\beta}e^{2i\pi\beta n/M}C_{\alpha, \beta+\alpha}.
\end{equation}
Following the definition of $V_i$ we define the quantities
\begin{equation}
\label{defvialpha}
V_i(\alpha)=\ln|\Psi_i(\alpha)|^2-\mathbb{E}\left\{\ln|\Psi_i(\alpha)|^2\right\}.
\end{equation}
Expanding the logarithmic functions in \eqref{defvialpha} we find that at the lowest order the covariance structure for a fixed vector reads
\begin{equation}
 \mathbb{E}\left\{ V_i(\alpha)V_j(\alpha)\right\}=\mathbb{E}\left\{ A_i(\alpha)A_j(\alpha)\right\}-\mathbb{E}\left\{ A_i(\alpha)\right\}\mathbb{E}\left\{ A_j(\alpha)\right\}
\end{equation}
where we have defined
\begin{equation}
A_i(\alpha)=Q_i(\alpha)+Q_i^*(\alpha).
\end{equation}
Since we are interested in covariances of the form $\langle V_kV_{k+r}\rangle$ averaged over the value of $k$, we need to calculate terms of the form
\begin{equation}
\frac{1}{M}\sum_k Q_k(\alpha)Q_{k+r}(\alpha)=\sum_{\beta}e^{-\frac{2i\pi}{M}\beta r}C_{\alpha, \alpha+\beta}C_{\alpha, \alpha-\beta}.
\label{sommeQQ}
\end{equation}
We are interested in quantities averaged over all eigenvectors and random phases: the average $\mathbb{E}\{\ldots\}$ amounts to taking the expectation value with respect to $\alpha$ and an integral over the phases $\Phi_j$.

From Eqs.~\eqref{M1RS}--\eqref{cabb}, the explicit expression for $C_{\alpha, \beta+\alpha}$ is found to be
\begin{eqnarray}
\label{caba}
C_{\alpha, \beta+\alpha}&=&\frac{i}{2M}\sum_{mn}\frac{t_{m-n+1}}{\sin(\pi\beta/M)}
\exp\left[i(n-m-1)(\frac{2\pi\alpha}{M}+\tilde{\Phi})\right.\nonumber\\
&-&\left.\frac{2i\pi}{M}(m+\frac12)\beta-i\sum_{j=0}^{n-1}\Phi_j+i\sum_{j=0}^{m}\Phi_j\right],
\end{eqnarray}
with
\begin{equation}
\label{deft}
t_x=\frac{\pi\epsilon}{M}\frac{e^{-i\pi x/M}}{\sin\pi x/M}\quad\textrm{if $x\neq 0$,}\qquad \mbox{and} \quad t_x=0\quad \textrm{ otherwise.}
\end{equation}
 The only dependence on $\alpha$ in \eqref{caba} is via $\exp[2i\pi (n-m-1)\alpha/M]$. Then the averaging over $\alpha$ in \eqref{sommeQQ} yields a coefficient
\begin{equation}
\label{sommealpha}
\frac{1}{M}\sum_{\alpha}e^{\frac{2i\pi}{M}(n-m-1+n'-m'-1)\alpha}=\delta_{n-m+n'-m'-2},
\end{equation}
where non-primed indices correspond to the sum featuring in $C_{\alpha, \beta+\alpha}$ and primed ones to the sum in $C_{\alpha, \alpha-\beta}$. Because of the term \eqref{sommealpha}, $n-m+n'-m'-2=p M$ for some integer $p$, so that the contribution of $\tilde{\Phi}$ in $C_{\alpha, \beta+\alpha}C_{\alpha, \alpha-\beta}$ is of the form $\exp[i p M \tilde{\Phi}]=\exp[i p \sum_j\Phi_j]$. Thus, the averaging of \eqref{sommeQQ} over random phases contains a coefficient
\begin{equation}
\label{sommeangles}
 \mathbb{E}\left\{\exp\left(-i\sum_{j=0}^{n-1}\Phi_j+i\sum_{j=0}^{m}\Phi_j-i\sum_{j=0}^{n'-1}\Phi_j+i\sum_{j=0}^{m'}\Phi_j+i p \sum_{j=0}^{M-1}\Phi_j\right)\right\}_{\Phi}.
\end{equation}
Obviously only terms where all phases in the exponent cancel each other can survive the average. At the same time the term $t_{m-n+1}$ in \eqref{caba} implies that contributions with $m-n+1=0$ or $m'-n'+1=0$ must vanish. [Note that in particular this means that the terms $\mathbb{E}\{ A_i(\alpha)\}$ vanish.] The only remaining possibility in \eqref{sommeangles} is to have simultaneously $m=n'-1$ and $m'=n-1$. This yields
\begin{equation}
\label{CCrandom}
\mathbb{E}\left\{ C_{\alpha, \alpha+\beta}C_{\alpha, \alpha-\beta}\right\}_{\alpha, \Phi}=
\frac{1}{4M^2}\sum_{mn}\frac{|t_{m-n+1}|^2e^{-\frac{2i\pi}{M}(m-n+1)\beta}}{\sin^2(\pi\beta/M)}.
\end{equation}
Changing variables $m-n+1=x$ and summing over $\beta$ we get from \eqref{sommeQQ} and \eqref{deft}
\begin{equation}
\mathbb{E}\left\{\frac{1}{M}\sum_k Q_k(\alpha)Q_{k+r}(\alpha)\right\}_{\alpha, \Phi}=
-\frac{\pi^2\epsilon^2}{4M^3}\sum_{x,\beta}\frac{e^{-\frac{2i\pi}{M}x\beta-\frac{2i\pi}{M}\beta r}}{\sin^2(\pi x/M)\sin^2(\pi\beta/M)}.
\end{equation}
In a similar way, for terms of the form $Q Q^*$, Eq.~\eqref{sommeQQ} becomes
\begin{equation}
\frac{1}{M}\sum_k Q_k(\alpha)Q_{k+r}(\alpha)^*=\sum_{\beta}e^{-\frac{2i\pi}{M}\beta r}|C_{\alpha, \alpha+\beta}|^2.
\label{sommeQQstar}
\end{equation}
Averaging $C_{\alpha, \alpha+\beta}C^{*}_{\alpha, \alpha+\beta}$ over $\alpha$ yields, instead of \eqref{sommealpha}, a coefficient $\delta_{n-m-n'+m'}$, and the average over $\Phi_j$ then yields the condition $n=n'$ and $m=m'$, so that
\begin{equation}
\label{big1}
\mathbb{E}\left\{ C_{\alpha, \alpha+\beta}C^*_{\alpha, \alpha+\beta}\right\}_{\alpha, \Phi}=
\frac{1}{4M^2}\sum_{mn}\frac{|t_{m-n+1}|^2}{\sin^2(\pi\beta/M)}
\end{equation}
and
\begin{equation}
\label{big2}
\mathbb{E}\left\{\frac{1}{M}\sum_nQ_n(\alpha)Q_n(\alpha)^*\right\}_{\alpha, \Phi}\!\!\!\!
=\frac{\pi^2\epsilon^2}{4M^3}\sum_{x,\beta}\frac{e^{-\frac{2i\pi}{M}\beta r}}{\sin^2(\pi x/M)\sin^2(\pi\beta/M)}.
\end{equation}
Putting together Eqs.~\eqref{big1} and \eqref{big2}, we get
\begin{equation}
\fl\mathbb{E}\left\{\frac{1}{M}\sum_n(Q_n(\alpha)+Q_n^{*}(\alpha))(Q_{n+r}(\alpha)+Q_{n+r}^{*}(\alpha))
\right\}_{\alpha, \Phi}\!\!\!\!
=\frac{\pi^2\epsilon^2}{M^3}\sum_{x,\beta}\frac{\sin\frac{\pi\beta x}{M}\sin\frac{\pi\beta (x-2r)}{M}}{\sin^2\frac{\pi x}{M}\sin^2\frac{\pi\beta}{M}}.
\label{big3}
\end{equation}
One can evaluate the sum over $\beta$ as
\begin{equation}
\sum_{\beta}\frac{\sin\frac{\pi\beta x}{M}\sin\frac{\pi\beta (x-2r)}{M}}{\sin^2\frac{\pi\beta}{M}}=
\left|
\begin{array}{lc}
(x-2r)(M-x)&\qquad x\geq r\\
x(2r-x-M) &\qquad x\leq r\ .
\end{array}
\right.
\end{equation}
The covariance averaged over the eigenvectors $\alpha$, phases $\Phi$ and position $k$ finally reads
\begin{equation}
\label{corrfinal}
\fl\mathbb{E}\left\{ V_k(\alpha)V_{k+r}(\alpha)\right\}=\frac{\pi^2\epsilon^2}{M^3}
\left(\sum_{x<r}\frac{x(2r-x-M))}{\sin^2\frac{\pi x}{M}}+\sum_{x\geq r}\frac{(x-2r)(M-x)}{\sin^2\frac{\pi x}{M}}\right).
\end{equation}
This expression is invariant upon the change $r$ to $M-r$. We therefore consider only $r<M/2$. We are interested in the asymptotic behavior of \eqref{corrfinal} for $r=c M$ with fixed $c<1/2$. The long-distance behaviour of the covariance corresponds to extracting the asymptotics in terms of $\ln(|i-j|/M)$ when $i$ and $j$ are considered to be distant. This corresponds to the limit $c\to 1$, or equivalently, by symmetry of \eqref{corrfinal}, to $c\to 0$.
\begin{figure}[t!]
\begin{center}
\includegraphics*[width=.85\linewidth]{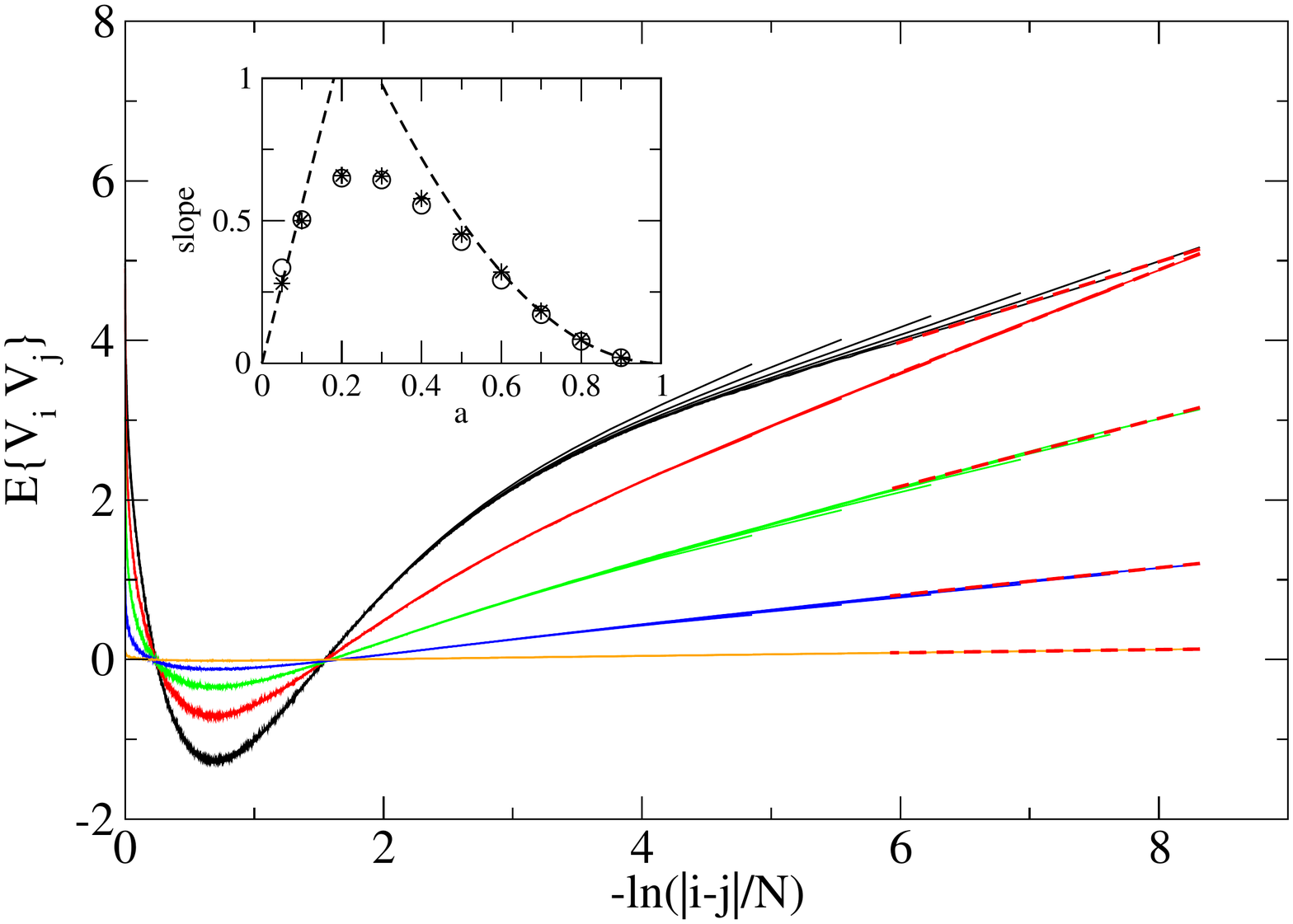}
\end{center}
\caption{Correlation function for Ruijsenaars model with $a=0.1$ (black), 0.3 (red), 0.5 (green), 0.7 (blue), 0.9 (orange). Various curves of the same color correspond to sizes from $2^7$ to $2^{12}$ (with $2^{29-2n}$ realizations of the random matrices) and come on top of each other. Dashed red lines are a linear fit (performed on the curve for $M=2^{12}$) on the end of the curve. Inset: slopes of these fits as a function of $a$ (circles), and second derivatives of $\tau_q$ at $q=0$ (stars) obtained from a quadratic fit of $\tau_q$ in the interval $q\in [-0.3,0.3]$. The dashed curves indicate the perturbation-theory value $-\tau''_q|_{q=0}=4 a \ln 4$ for $a\simeq 0$, obtained from \eqref{dqstrong}, and $-\tau''_q|_{q=0}=-(1-a)^2$ for $a\simeq 1$, obtained from \eqref{dqweak}. Curves for correlations corresponding to values $a=0.05$, $0.2$, $0.4$, $0.6$, $0.8$ are not shown in the main panel, only the slope of the corresponding fit is given in the inset for these values.
\label{correl_ruij}}
\end{figure}
For $c<\frac12$, the right-hand side of \eqref{corrfinal} can be rewritten
\begin{equation}
\label{corrfinal2}
\frac{\pi^2\epsilon^2}{M^3}
\left(\sum_{x=1}^{cM-1}\frac{-2x^2}{\sin^2\frac{\pi x}{M}}+\sum_{x=cM}^{(1-c)M}\frac{x(M(1-2c)-x)}{\sin^2\frac{\pi x}{M}}\right).
\end{equation}
The first term is a Riemann sum which converges to the integral
\begin{equation}
-2\pi^2\epsilon^2\int_{0}^{c}dy\frac{y^2}{\sin^2 \pi y},
\end{equation}
which is non-singular when $c\to 0$. The second term is another Riemann sum for the integral
\begin{equation}
I_c=\pi^2\epsilon^2\int_{c}^{1-c}dy\frac{y(1-2c-y)}{\sin^2\pi y},
\end{equation}
which in contrast diverges when $c\to 0$. We are interested in extracting the small-$c$ behaviour of the integral $I_c$. To this end one can employ the identity
\begin{equation}
\pi^2\int_{0}^{1}dy\left(\frac{y(1-y)}{\sin^2\pi y}-\frac{1}{\pi^2 y}-\frac{1}{\pi^2 (1-y)}\right)=2(1-\ln(2\pi)),
\end{equation}
which allows for the integral $I_c$ to be rewritten as
\begin{eqnarray}
\label{intic}
I_c&=&\pi^2\epsilon^2\int_{c}^{1-c}dy\left(\frac{y(1-y)}{\sin^2\pi y}-\frac{1}{\pi^2 y}-\frac{1}{\pi^2 (1-y)}\right)\\
&+&\epsilon^2\int_{c}^{1-c}dy\left(\frac{1}{y}+\frac{1}{1-y}\right)-2c\pi^2\epsilon^2\int_{c}^{1-c}dy\frac{y}{\sin^2\pi y}.\nonumber
\end{eqnarray}
The first term converges to a constant when $c\to 0$. The third term can be straightforwardly evaluated using
\begin{equation}
\pi^2 c\int_{c}^{1-c}dy\frac{y}{\sin^2\pi y}=\pi c \cot (\pi c),
\end{equation}
which goes to a finite value when $c\to 0$. The middle term in \eqref{intic} gives
\begin{equation}
\label{lastintegral}
\int_{c}^{1-c}dy\left(\frac{1}{y}+\frac{1}{1-y}\right)=2\ln\frac{1-c}{c}\sim_{c\to 0} -2\ln c,
\end{equation}
which results in the final asymptotic expression for the covariance
\begin{equation}
\label{correlfin}
\mathbb{E}\left\{ V_k(\alpha)V_{k+r}(\alpha)\right\}\sim -2\epsilon^2\ln \frac{r}{M},\qquad r\ll M.
\end{equation}
This behaviour linear in $-\ln(r/M)$ is illustrated by the right part of the curves in the main panel of Fig.~\ref{correl_ruij}, which correspond to $|i-j|\ll M$. The expression \eqref{correlfin} exactly coincides with the expression \eqref{8} in this regime. Indeed, as mentioned earlier, when the map parameter $a$ is close to 1 one has $\tau''_q|_{q=0}=-2(1-a)^2=-2\epsilon^2$ from Eq.~\eqref{dqweak}, and Eq.~\eqref{8} gives precisely \eqref{correlfin}, with a minus sign corresponding to the sign difference between $\zeta_q$ and $\tau_q$.

For arbitrary $a$, where perturbation theory does not apply, Eq.~\eqref{8} can be checked by extracting the second derivative of $\tau_q$ at 0 from a quadratic fit around zero (the second derivatives of $\tauqtyp$ yield the same values up to our numerical precision). In Fig.~\ref{correl_ruij} we show the covariance as a function of $\ln(|i-j|/M)$. We observe that the curves corresponding to different $M$ all come on top of each other. Extracting the slopes of the correlation at $|i-j|\ll M$ we get the circles in the inset of Fig.~\ref{correl_ruij}. They match the second derivatives, apart for $a\ll 1$, where the numerical evaluation of $\tauq$ is less reliable.

\section{Comparison with simpler models}
 Our numerical results for the extreme value statistics in the Ruijsenaars-Schneider ensemble are limited by the fact that numerically extracting the singularity spectrum around $\alpha_{-}$ is quite hard, given that precise computation of the average multifractal exponents of eigenvectors requires diagonalization of many matrix realizations.

\begin{figure}[t]
\begin{center}
\includegraphics*[width=.48\linewidth]{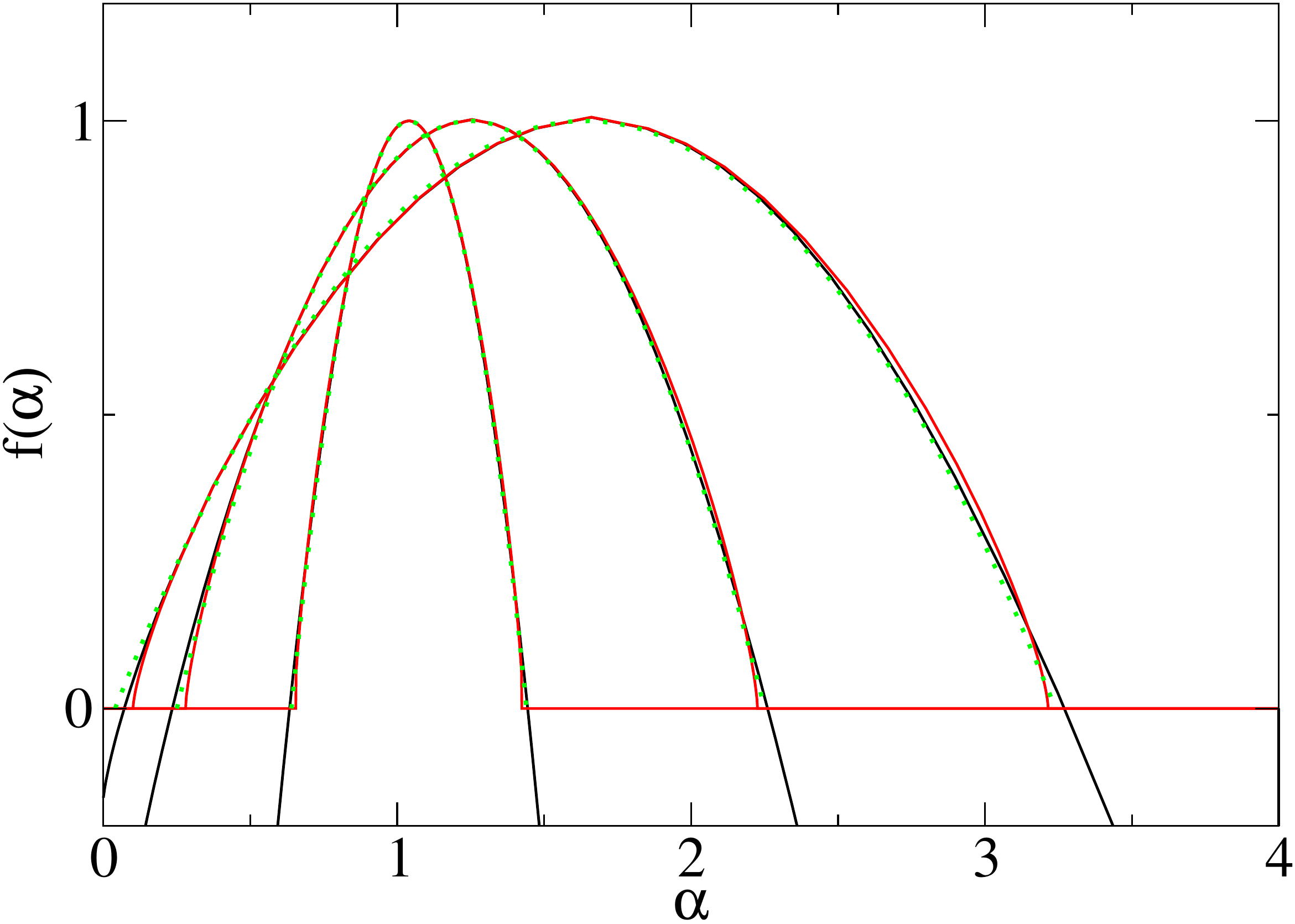}
\includegraphics*[width=.48\linewidth]{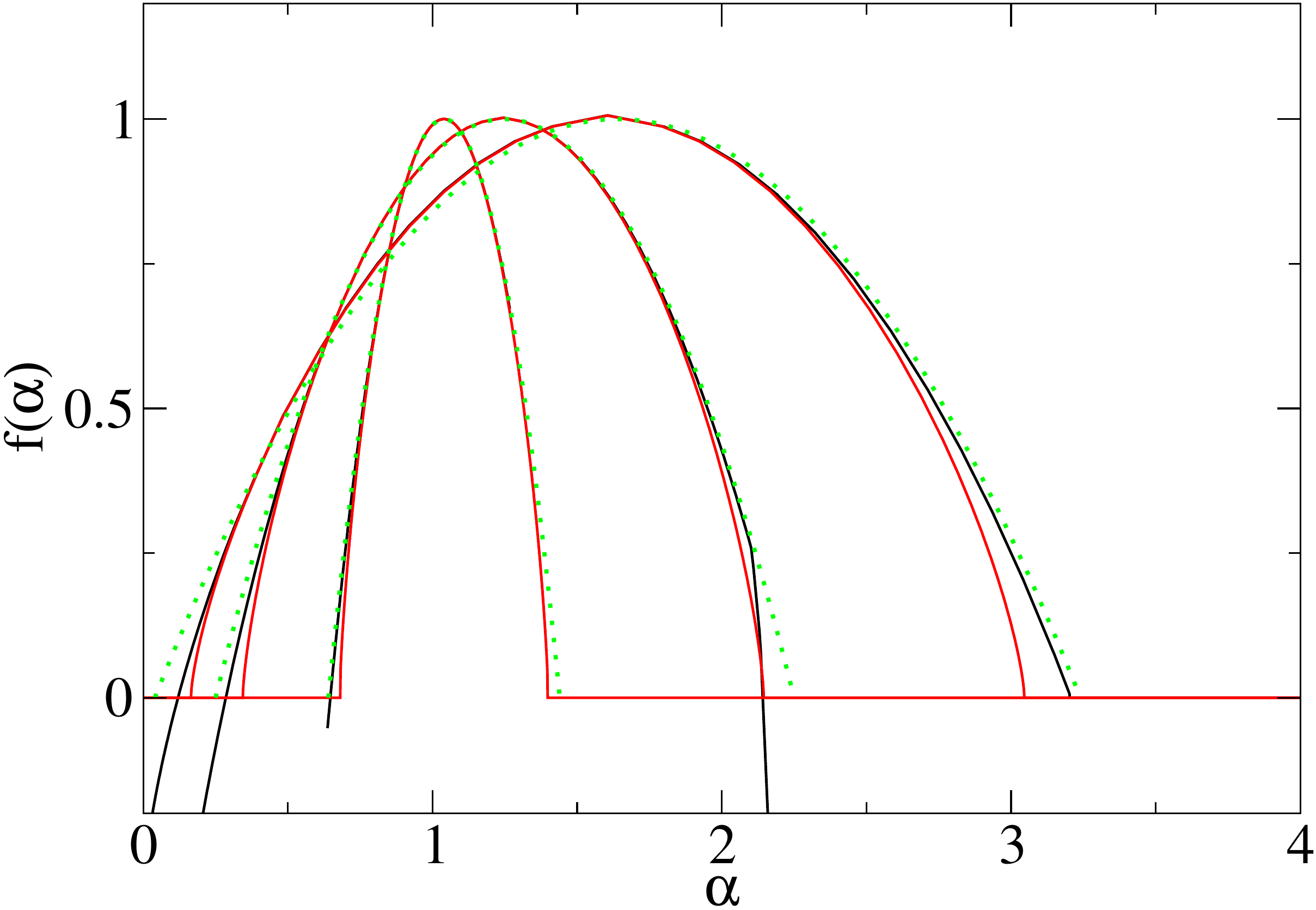}
\end{center}
\caption{Singularity spectrum $f(\alpha)$ (black) and $\ftyp(\alpha)$ (red/grey) for the REM (left) and DSM (right), from narrowest to widest $\beta=0.2, 0.5, 0.8$. The dotted green curve  corresponds to the analytic expression \eqref{falpha_rem}. Here multifractal exponents $\tau_q$ and $\tauqtyp$ are obtained by averaging over $2^{18}$ vector realizations at each size and fitting over the range $2^7-2^{12}$.\label{falrem}}
\end{figure}
\begin{figure}[t]
\begin{center}
\includegraphics*[width=0.48\linewidth]{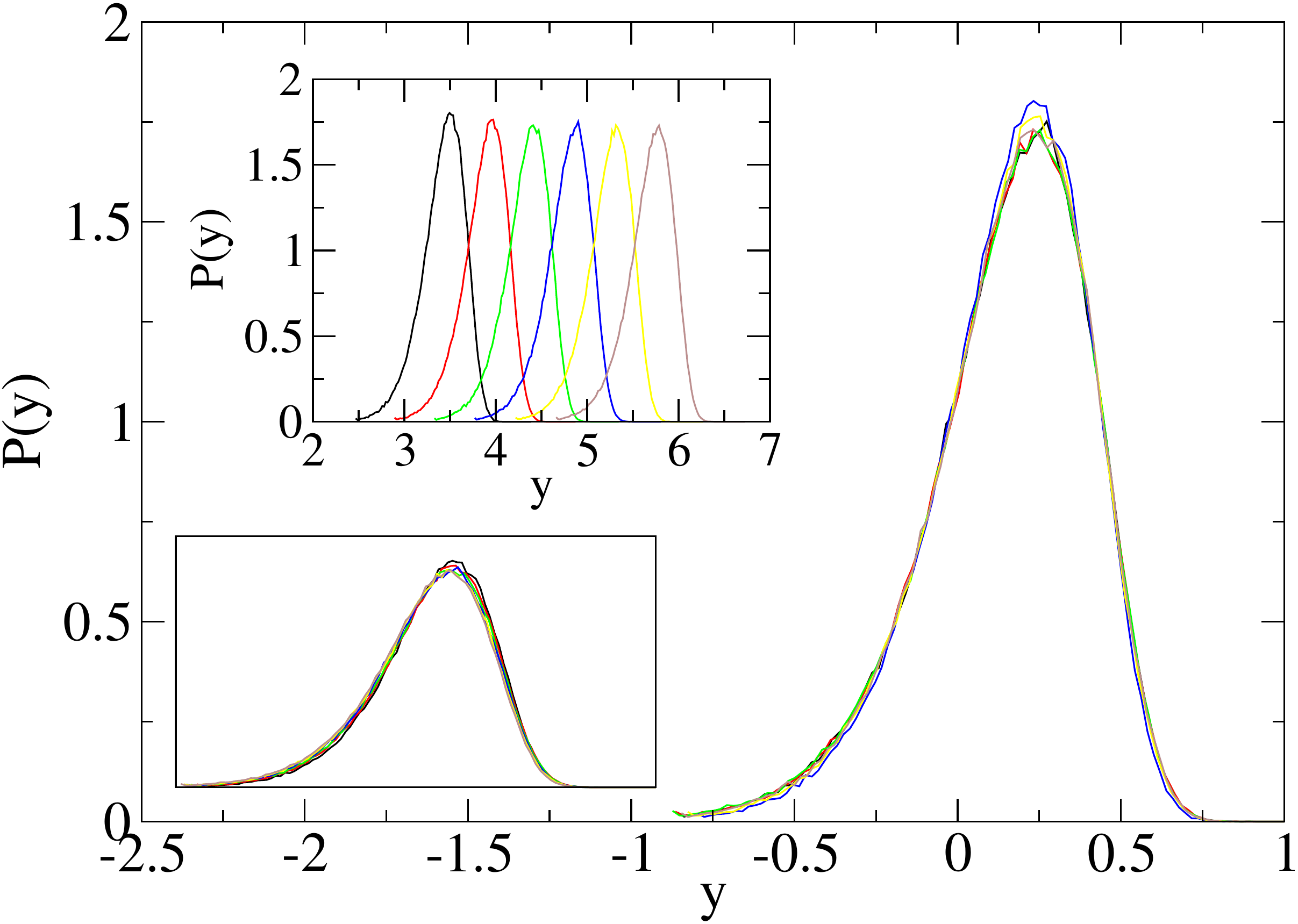}\includegraphics*[width=0.48\linewidth]{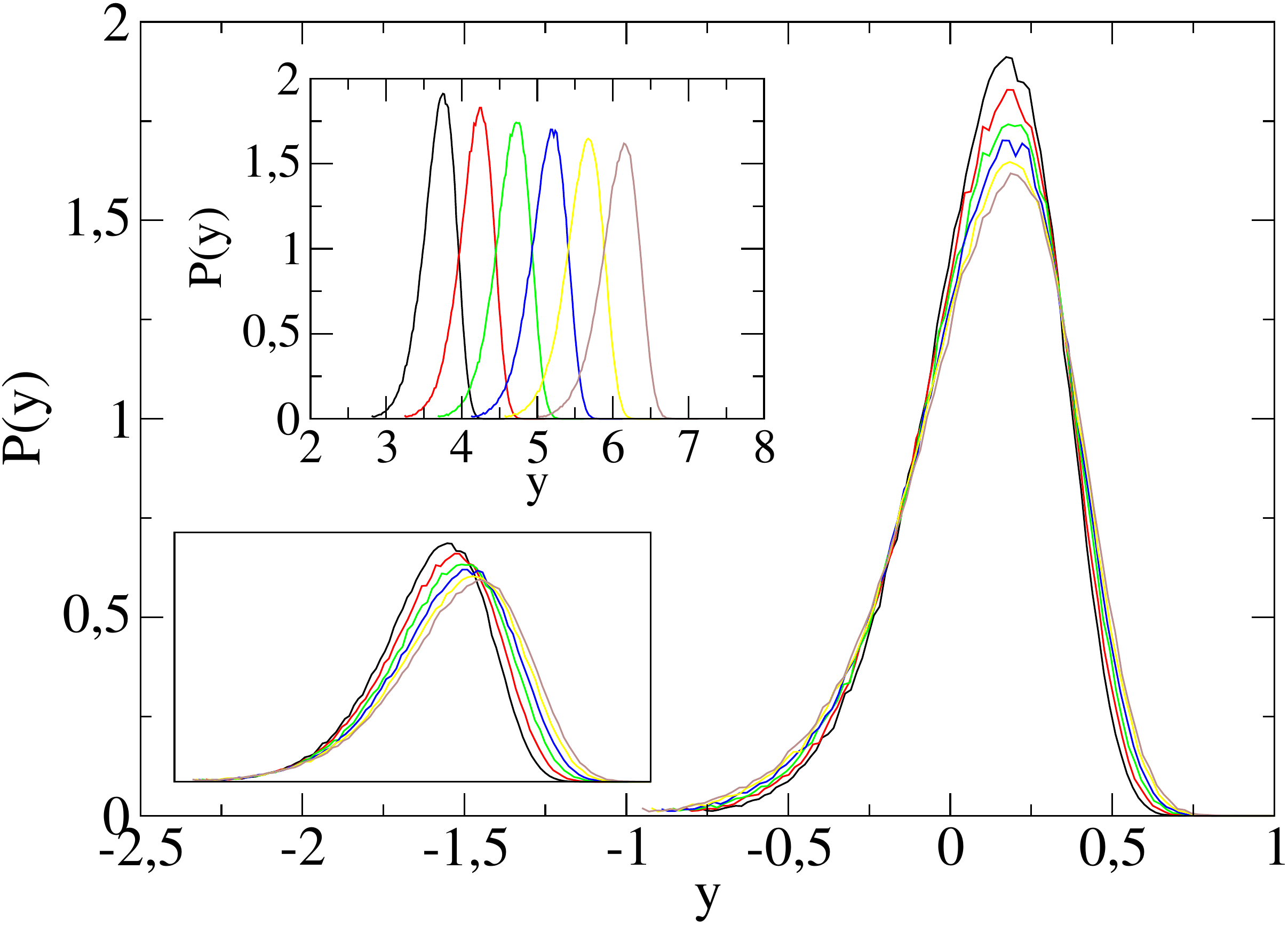}\\
\includegraphics*[width=0.48\linewidth]{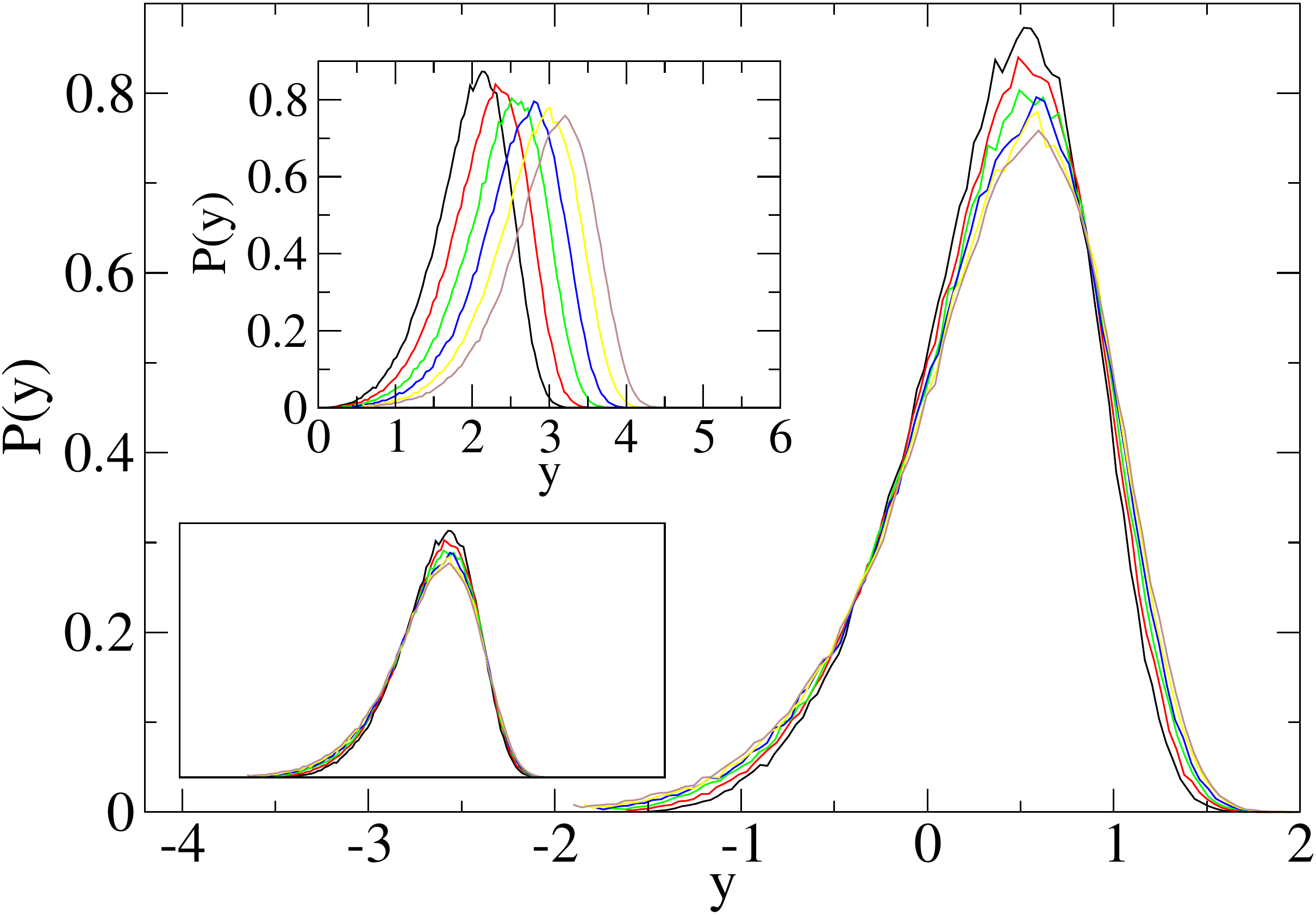}\includegraphics*[width=0.48\linewidth]{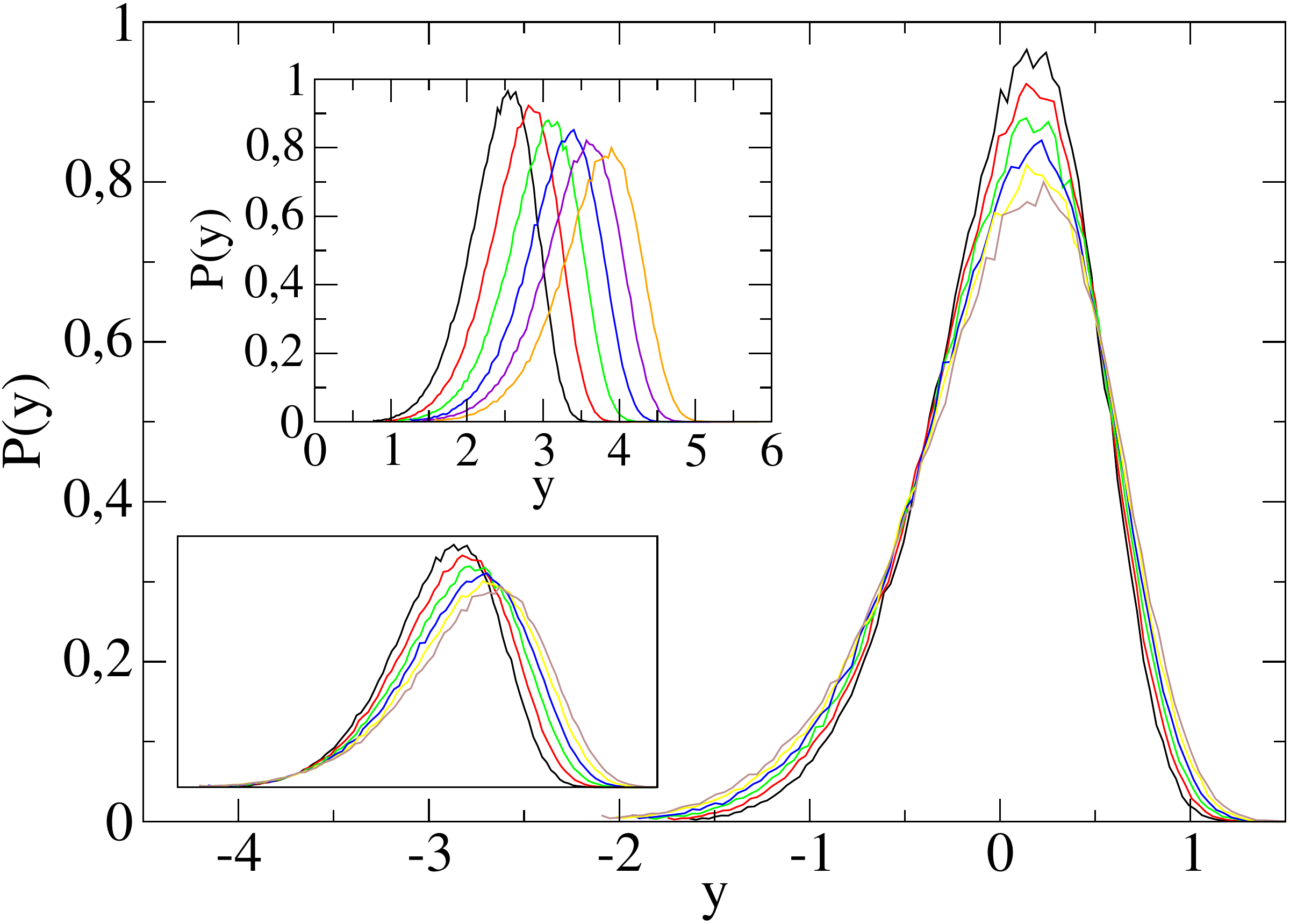}\\
\includegraphics*[width=0.48\linewidth]{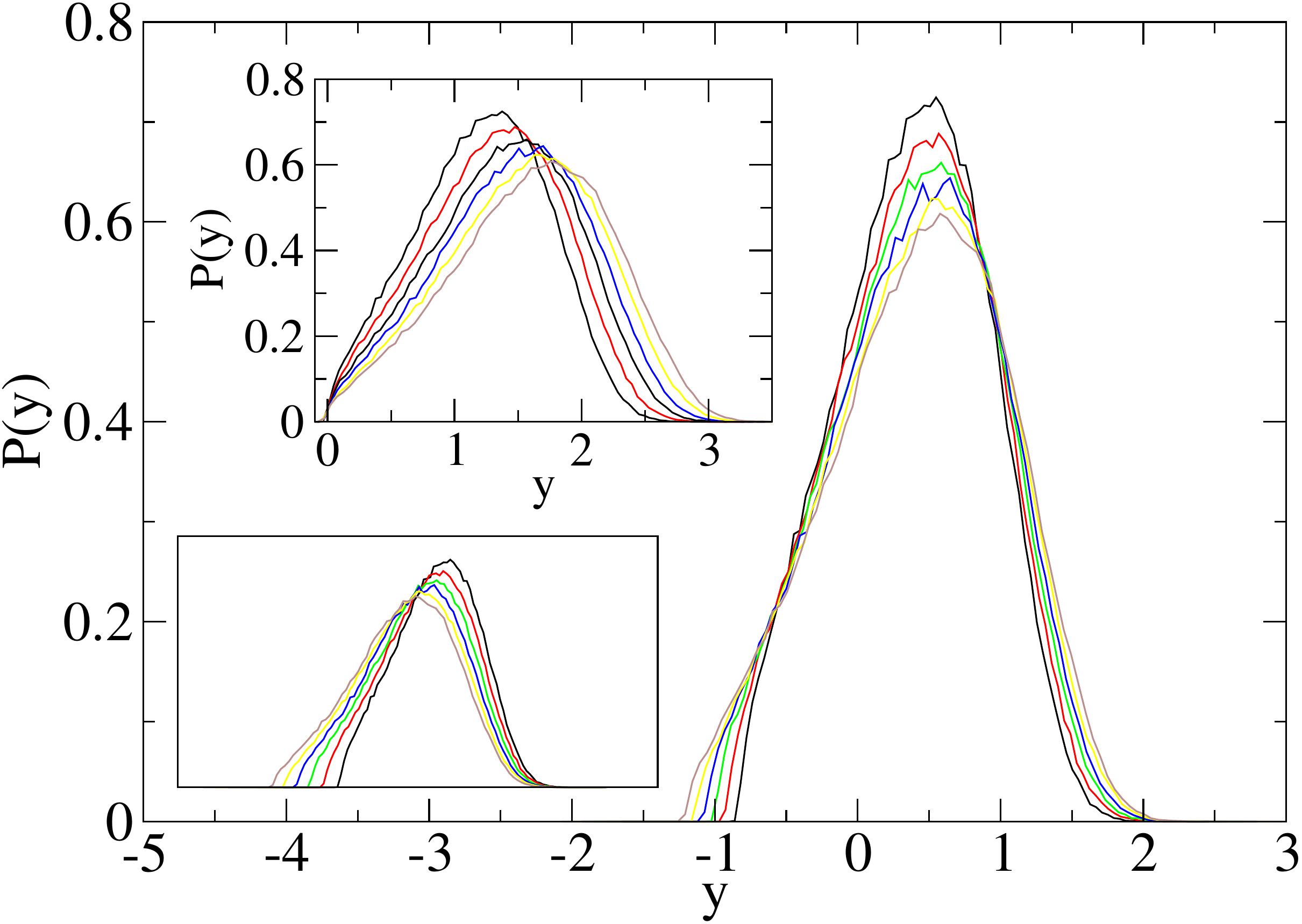}\includegraphics*[width=0.48\linewidth]{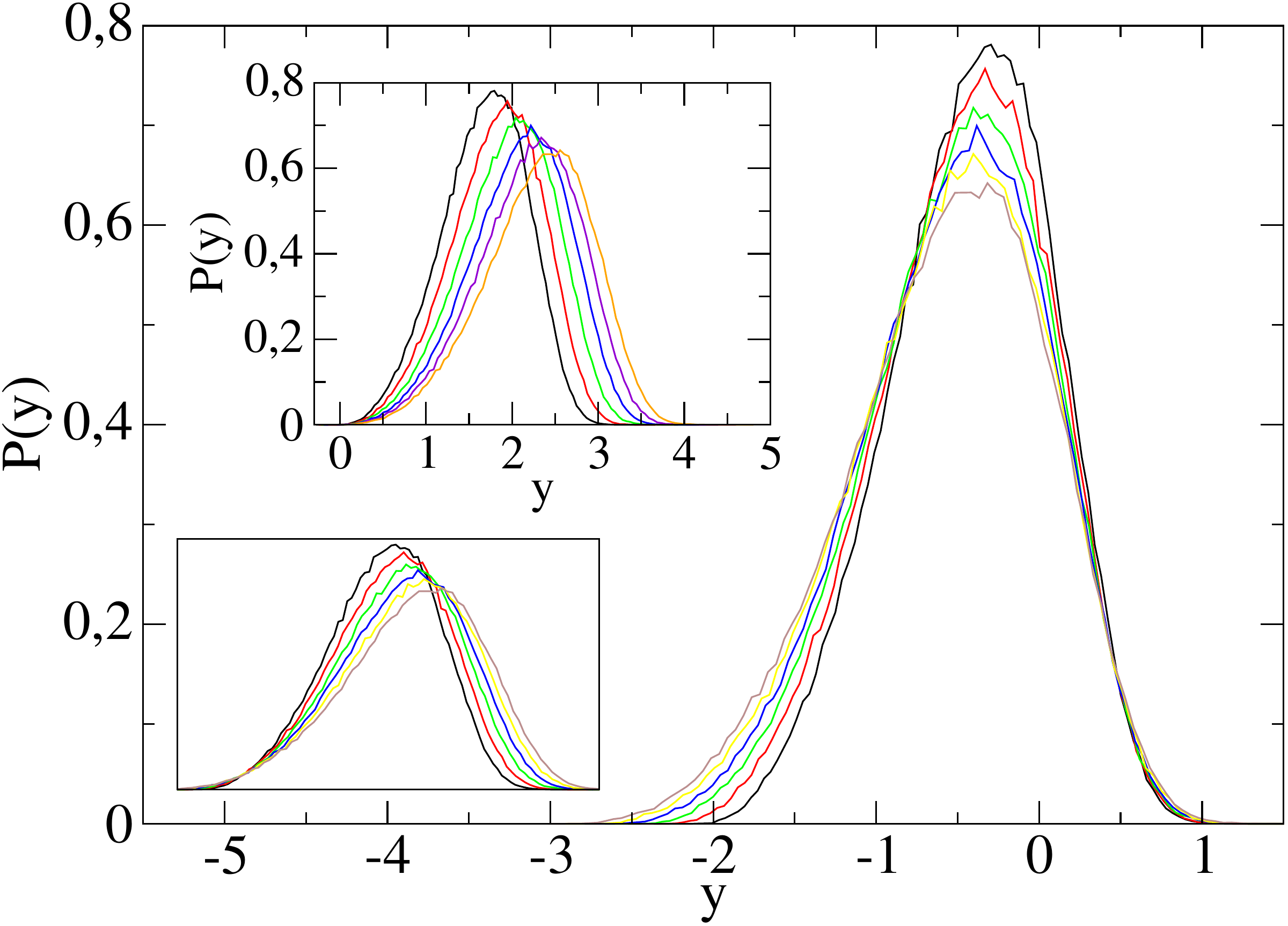}
\end{center}
\caption{Distribution of $y=-\ln p_m$ for REM (left column) and DSM (right column) for $\beta=0.2$, 0.5 and 0.8 and $M=2^{n}$ with $n=7$ to $12$ (from black to brown), data from $2^{18}$ realizations. Top inset: original data. Main panel: shift by $(1-\beta)^2\ln M+c\frac{\beta}{2}\ln\ln M$, with $c=1/2$ (REM) or $3/2$ (DSM). Lower inset: shift with $c=3/2$ (REM) or $1/2$ (DSM).
\label{pdemax}}.
\end{figure}
To put these results in perspective, we compared numerics for RSE  with two simple examples of ensembles of random vectors, where analytic expressions for the singularity spectrum exist. Both models are defined by $p_i=e^{\beta V_i}/Z(\beta)$ with $V_i$, $1\leq i\leq M$, random variables, and $Z(\beta)$ the normalization constant. The first model is the random energy model (REM) \cite{Derr81}, where $V_i$ are taken as independent and identically distributed Gaussian random variables with $\langle V_i\rangle=0$ and $\langle V_i^2\rangle=2\ln M$. In the second model, proposed by Derrida and Spohn (DSM) \cite{DS}, the $V_i$ are defined as a sum of independent Gaussian random variables as follows: we consider a binary tree of edges connecting the top vertex (the "root") with $M=2^n$ "leaves" (i.e. vertices at the finest hierarchy level with no outgoing edges), and attribute to each edge of the tree a centered Gaussian random variable with variance $\frac{2n}{n+1}\ln 2$. Then $V_i$ is the sum of all $n+1$ variables along the path going from the root to leaf $i$ of the tree. As a sum of independent Gaussian variables, they are also Gaussians with variance $2n\ln 2=2\ln M$.

It can be shown that for both models the singularity spectrum is given by
\begin{equation}
f(\alpha)=1-\frac{1}{4\beta^2}(1+\beta^2-\alpha)^2
\label{falpha_rem}
\end{equation}
for $\alpha\in[(1-\beta)^2,(1+\beta)^2]$ and zero outside this interval. However the expected behavior of the maxima of $p_i$ will differ because of logarithmic correlations in DSM which are absent in REM. Following Eq.~\eqref{16}, the maximum is expected to scale as
\begin{equation}
\label{maxpi}
-\ln p_m \simeq (1-\beta)^2\ln M+c\beta\ln\ln M,
\end{equation}
with $c=1/2$ for REM and $c=3/2$ for DSM. The values of $\alpha_{-}$ and $f'(\alpha_{-})$ are available analytically from \eqref{falpha_rem}, but again are difficult to extract numerically, especially for small $\beta$, as is illustrated in Fig.~\ref{falrem}, where Eq.~\eqref{falpha_rem} is compared against numerical data.

Equation \eqref{maxpi} is compared with numerical results in Fig.~\ref{pdemax}. It clearly shows that the $3/2$ coefficient is required for DSM; this is less clear for REM.

\section{Conclusion}
In the present paper we have reviewed properties of extreme values in multifractal patterns. In particular we have underlined the relationship between logarithmically correlated random processes and disorder-generated multifractals: the logarithm of a disorder-generated multifractal appears as a log-correlated random field. This connection was exemplified on an ensemble of random matrices with multifractal eigenvectors, the Ruijsenaars-Schneider ensemble. We showed analytically, in the perturbation-theoretic regime of weak multifractality, that the logarithm of eigenvector intensities displays logarithmic correlations which are related to multifractal dimensions precisely in the way expected for a generic logarithmically correlated random process.

This connection between these two types of models allows to draw a parallel between features of their extreme values. In $1/f$ noises, extreme values follow Eq.~\eqref{13}, while for the logarithm of multifractal patterns they follow the analogous relation Eq.~\eqref{16}. We investigated the validity of this latter relation in several models of disorder-generated multifractals, including the Ruijsenaars-Schneider ensemble and models with (DSM) or without (REM) logarithmic correlations. We checked that these relations are verified and indeed allow to predict the behaviour of extreme values from the knowledge of the multifractality spectrum.

\end{document}